\documentstyle[12pt]{article}
\begin{document}
\setlength{\unitlength}{1mm}

{\hfill   May 1998 }

{\hfill    Alberta-Thy 10-98}

{\hfill    hep-th/9806004} \vspace*{2cm} \\
\begin{center}
{\Large\bf Correlation functions of boundary field theory from bulk
Green's functions and phases in the boundary theory}
\end{center}
\begin{center}
Sergey N.~Solodukhin\footnote{e-mail: sergey@phys.ualberta.ca}
\end{center}
\begin{center}
{\it Department of Physics, University of Alberta, Edmonton, Alberta, T6G 2J1, 
Canada}  
\end{center}
\vspace*{1cm}
\begin{abstract} 
In the context of the bulk-boundary correspondence we study
the correlation functions arising on a boundary for different
types of boundary conditions.  The most general condition is
the mixed one interpolating between the Neumann and Dirichlet
conditions.  We obtain the general expressions for the 
correlators on a boundary in terms
of Green's function in the bulk for the Dirichlet, Neumann and mixed
boundary conditions and establish the relations between the 
correlation functions. As an instructive example
we explicitly obtain the boundary correlators corresponding 
to the mixed condition on a plane boundary $R^d$
of a domain in flat space $R^{d+1}$.
The phases of the boundary theory with correlators of the Neumann
and Dirichlet types are determined. The boundary correlation functions
on sphere $S^d$ are calculated for the Dirichlet and Neumann conditions
in two important
cases: when sphere
is a boundary of a domain 
in flat space $R^{d+1}$ and  when it is a boundary  
at infinity  of Anti-De Sitter space $AdS_{d+1}$.
For massless in the bulk theory the Neumann correlator on the boundary
of AdS space is shown to have universal logarithmic behavior in all
AdS spaces. In the massive case it is found to be finite at
the coinciding points. We argue that the Neumann correlator may have a dual
two-dimensional description.
The structure of the correlators obtained, their conformal nature 
and some recurrent relations are 
analyzed. We identify the Dirichlet and Neumann
phases living on the boundary of AdS space and discuss 
their evolution when
the location of the boundary changes from infinity to the 
center of the AdS space.

\end{abstract}
\begin{center}
{\it PACS number(s): 04.70.Dy, 04.62.+v}
\end{center}
\vskip 1cm
\newpage
\section{Introduction}
It was recently proposed in \cite{2}, \cite{3} that there is
a correspondence between theories defined in the bulk and on the boundary.
Particularly, the quantum correlation functions of the boundary theory
are expressed in terms of classical Green's functions in the bulk. 
This correspondence
was demonstrated in \cite{2}, \cite{3} for Anti-de Sitter (AdS) space
the spatial infinity of which plays the role of the boundary.
Green's function in the bulk arises in the boundary value problem with
 the Dirichlet condition
at infinity. The consideration of \cite{2}, \cite{3} is motivated
by the suggestion made in \cite{1} that the large $N$ limit of a superconformal Yang-Mill theory with gauge group $SU(N)$ in $d$ dimensions is governed
by supergravity on the product space $AdS_{d+1}\times\Sigma$ ($\Sigma$ is 
a compact manifold, often it is sphere $S^{d+1}$). Thus,  
one may hope to understand important features of QCD by just solving
the supergravity on the AdS space. Particularly, this may establish the long time suspected \cite{Pol} underlying string theory description  
of QCD in four dimensions since the supergravity under consideration is
what arises in the low-energy limit of 
type IIB superstring compactified on $AdS_5\times \Sigma$.
The works \cite{1}-\cite{3} initiated a flow of papers  developing further
the bulk-boundary (usually referred as CFT/AdS) 
correspondence \cite{4}-\cite{x2}.

The classical and quantum fields on AdS spaces are studied 
for long time \cite{x3}, \cite{x4}. This study is, in particular,
relevant to the physics of three-dimensional (BTZ) black holes 
( see \cite{19'} and references therein)
and to the extreme limit of higher-dimensional black holes \cite{19}.
That the AdS space has a boundary at infinity means that some 
condition should be imposed there. In the context of supergravity the
analysis made in \cite{x4} shows that there are two types of conditions
which preserve the supersymmetry: when one fixes the field function
$u$ (the Dirichlet type condition) or its normal derivative $\partial_n u$
(the Neumann type condition) at infinity of AdS space. 

In the present paper we investigate in a systematic way the correlation functions
arising on a boundary for different types of boundary conditions.
One of the starting points of our study is a simple observation
that the bulk-boundary correspondence is not a feature of  AdS spaces only
but is a general phenomenon. In the Euclidean version it arises
in an elliptic boundary value problem on arbitrary manifold with a boundary.
Different boundary conditions describe different theories on the boundary
or, better to say, different phases of the boundary theory.
The most  general condition one may impose  is the  mixed one 
when the
combination ($\partial_n u-hu$) 
is fixed on the boundary. Changing $h$ from zero to infinity it interpolates 
between the Neumann and Dirichlet conditions. Thus, in terms of the coupling
$h$ the Neumann phase appears on the boundary in the weak coupling
regime while the Dirichlet phase corresponds to the strong
coupling.
Not for any manifold the  elliptic boundary problem for
 the mixed condition can be solved explicitly. As a simple example when it can be done we
consider a domain of flat space $R^{d+1}$ with plane boundary $R^d$.
The boundary correlators corresponding to the mixed boundary condition
interpolate (when $h$ varies)  between the correlators arising
in the Neumann and Dirichlet problems. For a finite $h$ both phases
(with the Neumann and Dirichlet type correlators) present on the
boundary. 
When either boundary or manifold itself  (or both)
is curved the general picture remains the same
though the solving the elliptic problem becomes technically more 
difficult.
Therefore, one can solve the Neumann and Dirichlet
problems first and then apply the dimensional arguments to get                 
the structure of the correlation functions corresponding to the mixed boundary
condition in the regimes of weak and strong coupling $h$.
Proceeding this way, we calculate the Neumann and Dirichlet correlators
arising on sphere $S^d$ in two important cases: when $S^d$ is a boundary
of a domain in flat space $R^{d+1}$ and when it is a boundary at infinity of
space $AdS_{d+1}$.
These cases are actually related. They are the limiting points in the family of
boundary problems on a domain of AdS space when the boundary is shrinking
from infinity to the center of the space. 

Our results  for the Dirichlet phase at infinity of AdS space are 
in agreement with the consideration of \cite{2}, \cite{3}. 
What concerns the Neumann
phase, we find that it has quite remarkable feature: 
in the massless case the corresponding
boundary
correlation function is logarithmic for all AdS spaces.
We argue that there is a dual two-dimensional description
of this phase. In the context of the Yang-Mills theory
on the boundary of AdS space the Neumann phase may describe
that regime of the theory where the string (two-dimensional) 
nature of QCD manifests
the most. The mass in the bulk plays the role of a regulator in the
Neumann phase: the corresponding correlator becomes finite at
the coinciding points.

This paper is organized as follows. In the next Section we give a 
general consideration of the boundary-bulk correspondence
and obtain the expressions for the correlators on the boundary in terms
of Green's function in the bulk for the Dirichlet, Neumann and mixed
boundary conditions. We establish also the relations between
Green's functions corresponding to these conditions. The boundary value
problem with the mixed condition on the plane boundary $R^d$ of a
domain of flat space $R^{d+1}$ is explicitly solved and the corresponding
boundary correlators are analyzed in Section 3.
In Sections 4 and 5  we study th correlation functions
(of the Dirichlet and Neumann type)
arising on sphere $S^d$ considered as a boundary of a domain in flat space $R^{d+1}$ and space $AdS_{d+1}$ respectively.
The structure of the correlators obtained, their conformal nature 
and some recurrent relations are 
analyzed in Section 6.  In Section 7 we discuss the Dirichlet and Neumann
phases living on the boundary of AdS space and discuss their deformation in the limit when
the location of the boundary changes from infinity to the center of the AdS space. The presence of black hole inside AdS space is shown to affect the boundary theory in this limit. Throughout the paper we consider only Euclidean
version of the boundary-bulk correspondence.

\setcounter{equation}0
\section{Preliminary}
\setcounter{equation}0
Our starting point is the action
\begin{eqnarray}
W={1\over 2}\int_{\cal M} (\nabla u)^2
\label{1}
\end{eqnarray}
for the scalar field $u$ on Euclidean manifold $\cal M$ with boundary
$\cal B$. Varying (\ref{1}) with respect to $u$ we arrive at the Laplace equation
\begin{eqnarray}
\Box u=0~~.
\label{2}
\end{eqnarray}
The solving of this elliptic problem requires imposing on the function
$u$ some condition on the boundary $\cal B$. The minimal condition which can be
imposed is determined by considering the  term 
$\delta W_{\cal B}=\int_{\cal B}\partial_n u \delta u$ arising 
on the boundary under variation of the action (\ref{1}),
$\partial_n=n^\mu\partial_\mu$ is derivative with respect
to outer normal $n$  to the boundary $\cal B$. The term $\delta W_{\cal B}$
vanishes in two cases: i) if $\partial_n u=0$ on 
$\cal B$ or ii) value of $u$ is fixed on $\cal B$, particularly we may put
$u|_{\cal B}=0$. In fact, either of  these conditions is necessary to impose in
order to have the Laplace operator $\Box$ self-adjoint on the manifold
$\cal M$. Considering perturbation of these conditions we find
\begin{eqnarray}
\partial_n u|_{\cal B}=g
\label{3}
\end{eqnarray}
and 
\begin{eqnarray}
u|_{\cal B}=f
\label{4}
\end{eqnarray}
which are known respectively as 
the Neumann and Dirichlet boundary value problems, $g$ and $f$ are some
functions on the boundary. The function $g$ is not arbitrary.
It must satisfy
the condition
\begin{eqnarray}
\int_{\cal B}g=0~~.
\label{5}
\end{eqnarray}

The standard way to solve an elliptic boundary-value problem
is to apply Green's formula
\begin{eqnarray}
u(M)=\int_{\cal B}~ \left( G(M,P)\partial_n u(P)-
\partial_n G(M,P)u(P) \right) d\Sigma_P~~,
\label{6}
\end{eqnarray}
where $G(M,P)$ is source function (Green's function) defined as a solution
of the equation (\ref{2}) which has a singularity
at the point $M=P$. When point $M$ approaches point $P$ we have 
\begin{eqnarray}
&&G(M,P)\simeq {1\over 2\pi}\ln {1\over \sigma (M,P)}~,~~ if~ dim({\cal M})=2
\nonumber \\
&&G(M,P)\simeq {1\over (d-1)}{1\over \Sigma_d}{1\over \sigma^{d-1}(M,P)}~,~~
if~dim({\cal M})=d+1>2~~,
\label{7}
\end{eqnarray}
where $\sigma(M,P)$ is geodesical distance between  the points,
$\Sigma_d=\int_{S^d}1={2\pi^{d+1\over 2}\over \Gamma({d+1\over 2})}$
is area of $d$-dimensional sphere of unit radius.

Additionally on should impose the boundary condition 
\begin{eqnarray}
G(M,P)|_{P\in {\cal B}}=0
\label{8}
\end{eqnarray}
for the Dirichlet problem
and
\begin{eqnarray}
\partial_n G(M,P)|_{P\in {\cal B}}=0
\label{9}
\end{eqnarray}
for the Neumann problem. Then the solution of the elliptic boundary value
problem takes simple integral form:
\begin{eqnarray}
&&u_D=-\int_{\cal B}\partial_n G_D(M,P)f(P)d\Sigma_P ~~\nonumber \\
&&u_N=\int_{\cal B}G_N(M,P)g(P)d\Sigma_P+constant
\label{10}
\end{eqnarray}
where $D(N)$ refers to the Dirichlet (Neumann) case
respectively. In the Neumann case the solution is determined up
an irrelevant constant. Values of the harmonic function $u$ inside the 
manifold $\cal M$, thus, are completely determined by the boundary
conditions on $\cal B$.

This purely classical and elementary (actually, taken from the 
students text books
\cite{14}, \cite{15})
consideration of the boundary value problem
occurs to be important in the quantum theory. Indeed, a quantum state
of field $u$ on manifold $\cal M$ with boundary $\cal B$ is determined
by fixing the boundary condition (\ref{3}) or (\ref{4})
and considering the functional integral over all $u$ approaching the fixed
values at the boundary:
\begin{eqnarray}
&&\Psi_D[f,{\cal B}]=\int_{u|_{\cal B}=f} {\cal D} u e^{-W[u]} \nonumber \\
&&\Psi_N[g, {\cal B}]=\int_{\partial_nu|_{\cal B}=g}{\cal D} u e^{-W[u]}
\label{11}
\end{eqnarray}
Since we have different types of the boundary conditions there are
different quantum states $\Psi_D$ and $\Psi_N$.

For a free field described by the action (\ref{1}) it is easy to find 
how the quantum state $\Psi_{N(D)}$ depends on the condition on the
boundary. 
Indeed, arbitrary field $u$ in the integral (\ref{11})
defying D(N)-state can be represented in the form
$$
u=u_{D(N)}+u^q_{D(N)}~~,
$$
where $u_D$ $(u_N)$ is classical solution (\ref{10})
and $u^q_D$ $(u^q_N)$ is a ``quantum'' field with zero boundary condition:
$$
\partial_n u^q_N|_{\cal B}=0~,~~u^q_D|_{\cal B}=0~~.
$$
These conditions are what is necessary to impose on the boundary in order 
to the Laplace operator $\Box_D(\Box_N)$ be self-adjoint. 
The functional integration in
(\ref{11}) then can be performed and the result reads as follows
\begin{eqnarray}
&&\Psi_D[f, {\cal B}]=e^{-W[u_D]}~det^{-1/2}\Box_D \nonumber \\
&&\Psi_N[g, {\cal B}]=e^{-W[u_N]}~det^{-1/2}\Box_N
\label{12}
\end{eqnarray}
The dependence of the quantum state on the values of the quantum field on the
boundary, thus, is given by the classical action functional (\ref{1}) 
considered on the classical solution (\ref{10}):
\begin{eqnarray}
&&W={1\over 2}\int_{\cal B}u~\partial_n u~ d\Sigma
\nonumber \\
&&W[u_N]={1\over 2}\int\limits_{~~\cal B}\!\!\int g(P) G_N(P,P')g(P')d\Sigma_Pd\Sigma_P'
\nonumber \\
&&W[u_D]=-{1\over 2}\int\limits_{~~\cal B}\!\!\int f(P) \partial^{'}_n\partial_nG_N(P,P')f(P')d\Sigma_Pd\Sigma_P'
\label{13}
\end{eqnarray}
Note that due to the condition (\ref{5}) the adding of arbitrary constant to
$G_N$ does not change value of the functional $W[u_N]$ in (\ref{13}).

Since the quantum state (\ref{12}) is a functional of
values on the boundary $\cal B$ it is quite natural to relate
it with some field theory on $\cal B$. The proposal of 
\cite{2}, \cite{3} is to identify (\ref{12}) with the expectation value
\begin{equation}
\Psi_D[f,{\cal B}]=<e^{-\int_{\cal B}f {\cal O}_D}>_{\cal B}
\label{14}
\end{equation}
of an operator $\cal O_D$ arising in this boundary field theory.
Generalizing this proposal for the Neumann condition we have
\begin{equation}
\Psi_N[g,{\cal B}]=<e^{-\int_{\cal B}g {\cal O}_N}>_{\cal B}
~~.
\label{15}
\end{equation}
The functions $f$ and $g$ play the role of the external source
for the theory on the boundary $\cal B$. 
Note, that the right hand side of the equation (\ref{15})
does not change under shift the operator ${\cal O}_N$ on a constant:
${\cal O}_N\rightarrow {\cal O}_N+constant$. This is due to the condition (\ref{5}) for the Neumann source $g$.
Variation with respect to
the source standardly gives correlation functions
\begin{eqnarray}
&&<{\cal O}_D(P){\cal O}_D(P')>=-{1\over 2}\partial^{'}_n\partial_n
G_D(P,P') \nonumber \\
&&<{\cal O}_N(P){\cal O}_N(P')>={1\over 2}G_N(P,P')~~,
\label{16}
\end{eqnarray}
$P$ and $P'$ lying on the boundary $\cal B$,
which in our case are completely determined by the bulk Green's
functions.

An interesting generalization of the above construction is to
consider the so-called third boundary value problem
(or Robin type condition):
\begin{equation}
 \left(\partial_nu-hu\right)|_{\cal B}=g~~,
\label{17}
\end{equation}
where $h$ can in principle be some function on $\cal B$ but in the
simplest case it is just a constant. It is easy to see that the condition
(\ref{17}) is intermediate between the Dirichlet and Neumann 
conditions. Indeed,
in the limit $h\rightarrow 0$ eq.(\ref{17})
is just the Neumann condition while in the limit $h\rightarrow \infty$
we arrive at the Dirichlet boundary condition provided that $g=-hf$.

In order to accomplish the condition (\ref{17}) in the action
principle we need  add to (\ref{1}) some boundary term 
describing ``boundary interaction''. The total action is
\begin{equation}
W^{(h)}={1\over 2}\int_{\cal M}(\nabla u)^2-{1\over 2}\int_{\cal B}hu^2~~.
\label{18}
\end{equation}
Variation of (\ref{18}) with respect to $u$ gives us the ``minimal''
boundary condition \\ $\left(\partial_nu- hu\right)|_{\cal B}=0$. 
It determines
 $h$-dependent self-adjoint extension $\Box_{(h)}$ of the Laplace operator
on $\cal M$.

Defying the corresponding Green's function $G^{(h)}$ with the boundary 
condition
\begin{equation}
\left(\partial_n G^{(h)}(P,P')-h~G^{(h)}(P,P')\right)|_{P\in {\cal B}}=0
\label{19}
\end{equation}
and applying Green's formula (\ref{6}) we obtain the solution of the third
boundary value problem (\ref{17})
\begin{equation}
u^{(h)}=\int_{\cal B}G^{(h)}~g~~.
\label{20}
\end{equation}
The corresponding quantum state is defined as follows
\begin{equation}
\Psi_{(h)}[g,{\cal B}]=e^{-W^{(h)}[u^{(h)}]}~det^{-1/2}\Box_{(h)}~~,
\label{21}
\end{equation}
where
\begin{eqnarray}
&&W^{(h)}[u^{(h)}]={1\over 2}\int_{\cal B}u^{(h)}g \nonumber \\
&&={1\over 2}\int\limits_{~~\cal B}\!\!\int g(P) G^{(h)}(P,P')g(P')d\Sigma_Pd\Sigma_{P'}~~.
\label{22}
\end{eqnarray}
We see that the parameter $h$ can be viewed as strength of the boundary
interaction in (\ref{18}). The importance  of the states (\ref{21})
is that they give us a flow between $D-$ and $N-$ states.
Particularly, considering the boundary operator ${\cal O}_{(h)}(P)$ such that
$$
\Psi_{(h)}[g,{\cal B}]=<e^{-\int_{\cal B}g{\cal O}_{(h)}}>_{\cal B}
$$
we expect that the corresponding correlation functions
\begin{equation}
<{\cal O}_{(h)}(P){\cal O}_{(h)}(P')>={1\over 2}G_{(h)}(P,P')
\label{23}
\end{equation}
form a one-parametric family
interpolating between $N$- and $D$-correlators. 

Interestingly, the Neumann correlator arises in the weak coupling limit $(h\rightarrow 0)$ while the Dirichlet one appears in the
strong coupling regime $( h\rightarrow \infty)$.
Indeed, taking the limit $h\rightarrow 0$ in (\ref{19}) we obtain
the Neumann Green's function $G^{(h\rightarrow 0)}(P,P')=G_N(P,P')$
and the action (\ref{22}) coincides with the Neumann expression $W[u_N]$ (\ref{13}). On the other hand, we get the Dirichlet condition when taking the 
limit $h\rightarrow \infty$ in (\ref{19}), so we have $G^{(h\rightarrow \infty )}(P,P')
=G_D(P,P')$. Moreover, applying the normal derivative
$\partial_{n'}$ to eq.(\ref{19}) and considering $P'$ lying on the boundary as well we obtain the relation
$$
G^{(h)}(P,P')|_{P,P'\in {\cal B}}={1\over h^2} \partial_n\partial_{n'}
G^{(h)}(P,P')|_{P,P'\in {\cal B}}~~.
$$
Therefore, in the limit $h\rightarrow \infty$ we find that
$$
G^{(h\rightarrow \infty )}(P,P')|_{P,P'\in {\cal B}}={1\over h^2} \partial_n\partial_{n'}
G_D(P,P')|_{P,P'\in {\cal B}}~~
$$
and the action (\ref{22}) becomes minus the Dirichlet action
$W[u_D]$ (\ref{13}) if we substitute $g(P)=-hf(P)$. Hence, the correlator (\ref{23}) is the Neumann correlator for $h\rightarrow 0$ and is minus the Dirichlet
correlator for $h\rightarrow \infty$.

Not for any manifold $\cal M$ the $h$-problem can be solved explicitly.
In the next Section we give the explicit solution of the problem 
when $\cal M$ is flat space
and its boundary $\cal B$ is a  plane. However, it is always 
useful to remember
that $N$- and $D$-problems are just limiting cases of
more general class of problems.

It is worth noting that the boundary condition of the type (\ref{17}) arises 
in the important case of the scalar field coupled non-minimally 
to curvature ${\cal R}$ of $\cal M$:
\begin{equation}
W={1\over 2}\int_{\cal M}((\nabla u)^2-\xi {\cal R}u^2)-\int_{\cal B}
\xi {k} u^2~~,
\label{24}
\end{equation}
where $ k$ is extrinsic curvature of the boundary $\cal B$.
The adding of the boundary term in (\ref{24}) is necessary to have the 
well-defined
variation of $W$ (\ref{24}) with respect to metric \cite{18}. 
The natural boundary condition arising from
(\ref{24}) is the following
\begin{equation}
\left(\partial_n u-2\xi { k}u\right)|_{\cal B}=g
\label{25}
\end{equation}
Note that even if manifold $\cal M$ is flat (${\cal R}=0$) the non-minimal
coupling in (\ref{24}) still manifests if the boundary $\cal B$ has 
non-trivial extrinsic curvature. Note also that by adding a boundary term 
$a \int_{\cal B}u\partial_n u$ to (\ref{24})
one can change the relative coefficient in the left hand side of 
eq.(\ref{25}).

\bigskip

\section{Flat space with plane boundary, \\
Dirichlet-Neumann duality}
\setcounter{equation}0
In this Section we consider the case when the manifold $\cal M$ and its boundary $\cal B$ are $R^{d+1}$ and $R^{d}$ respectively.
In this case the consideration given in ithe previous Section is realized
explicitly.

Let $(x_1,...,x_d,z)$ be coordinates  in the space $R^{d+1}$.
We consider only a part of $R^{d+1}$ defined by condition $z\geq 0$.
So, the boundary ${\cal B}=R^d$ is the plane $z=0$ and the normal derivative
is $\partial_n=-\partial_z$. The fundamental
solution of the Laplace equation in $R^{d+1}$ is the function
$$
g_d\left(\rho^2+(z-z')^2\right)={1\over {(d-1)}}{1\over \Sigma_d}{1\over (
\rho^2+(z-z')^2)^{d-1 \over 2}}~~,~d>1
$$
$$
g_d\left(\rho^2+(z-z')^2\right)=-{1\over 4\pi}\ln {1\over (
\rho^2+(z-z')^2)}~~,~d=1
$$
where $\rho=\sqrt{(x_1-x_1')^2+...+(x_d-x_d')^2}$ is the distance
between the points on the plane $\cal B$. It is easy to construct Green's
function in both the Dirichlet and Neumann boundary value problems
\begin{equation}
G_{D(N)}^{(d)}=g_d\left(\rho^2+(z-z')^2\right)\mp g_d\left (\rho^2+(z+z')^2\right)~~,
\label{2.1}
\end{equation}
where sign $(-)$ stands for the Dirichlet problem and $(+)$ is for the Neumann
problem. The boundary conditions
$$
G_D|_{z=0}=0~~~, ~~~\partial_zG_N|_{z=0}=0 
$$
are obviously satisfied.

Green's function for the Robin type
condition (\ref{19}) can  be found explicitly \cite{16} as well.
Let us search it in the form
$$
G^{(h)}=G_N+v~~,
$$
where $v=v(x_1-x_1',...,x_d-x_d',z+z')$. Then from the boundary condition
$$
\left(\partial_zG^{(h)}-h~G^{(h)}\right)|_{z=0}=0
$$
we find the differential equation for the function $v=v(x_1-x_1',...,x_d-
x_d',\xi)$
$$
\partial_\xi v-h~v=2hg_d( \rho^2+\xi^2)~~.
$$
The solution reads
$$
v=-2h~e^{h\xi }\int_\xi^{+\infty}e^{-hs}g_d(\rho^2+s^2)ds~~,
$$
where the constant of integration has been chosen in such a way
that $v$ goes to zero when $\xi$ goes to infinity.
It is easy to check that the function $v$ is indeed a solution of the
Laplace equation. Green's function for the mixed type boundary value
problem then reads
\begin{equation}
G^{(h)}_{d}=
g_d\left(\rho^2+(z-z')^2\right)+ g_d\left(\rho^2+(z+z')^2\right)
-2h~e^{h(z+z') }\int_{(z+z')}^{+\infty}e^{-hs}g_d\left(\rho^2+s^2\right)ds~
\label{2.2}
\end{equation}
It is straightforward to check that in the limit $h\rightarrow 0$
Green's function $G^{(h)}$ indeed goes to the Neumann Green's function
$G_N$ (\ref{2.1}) as it was anticipated in the previous Section.
On the other hand in the limit of large positive $h$
we have
$$
\int_{(z+z')}^{+\infty}e^{-hs}g_d\left(\rho^2+s^2\right)\simeq
g_d\left(\rho^2+(z+z')^2\right)\int_{(z+z')}^{+\infty}e^{-hs}ds
$$
$$
={1\over h}e^{-h(z+z')} g_d\left(\rho^2+(z+z')^2\right)
$$
and the Dirichlet Green's function $G_{D}$ arises in (\ref{2.2}).
This is also in accord with the expectations of the previous Section.

The action functional calculated on the 
classical solution, thus, takes the form (\ref{22})
with Green's function $G^{(h)}$ in the form (\ref{2.2}). We find
the correlator induced on the boundary $\cal B$ 
\begin{eqnarray}
&&<{\cal O}^{(h)}(P){\cal O}^{(h)}(P')>=
K^d_{(h)}={1\over 2}G_{(h)}(P,P')~~, \nonumber \\
&&K^{d>1}_{(h)}={1\over (d-1)}{1\over \Sigma_d}\left( {1\over \rho^{d-1}}-
h\int_0^{+\infty}{e^{-hs} \over (\rho^2+s^2)^{d-1\over 2}}ds \right)~~,
\nonumber \\
&&K^{d=1}_{(h)}={1\over 4\pi}\left(- \ln \rho^2+h\int_0^{+\infty}e^{-hs}
\ln (\rho^2+s^2)ds \right)~~,
\label{2.3}
\end{eqnarray}
where $\rho$ is the distance between points $P$ and $P'$ on $R^d$.
Another representation  for the correlators (\ref{2.3}) is the following
\begin{equation}
K^d_{(h)}=-{1\over \Sigma_d}\partial_h\left( \int_0^{+\infty}
{e^{-hs} \over (\rho^2+s^2)^{d+1\over 2}}ds \right)~~.
\label{A}
\end{equation}

In the weak coupling limit $h\rightarrow 0$ the correlators $K^d_{(h)}$
go to the ones arising on the boundary $\cal B$ in the Neumann
problem 
\begin{eqnarray}
&&K_N^d=g_d(\rho^2)~~, \nonumber \\
&&K^{d>1}_N={1\over (d-1)}{1\over \Sigma_d}{1\over \rho^{d-1}}~~,~~K^{d=1}_{N}=
{1\over 2\pi}\ln \rho~~.
\label{N}
\end{eqnarray}
 In the strong coupling limit $h\rightarrow +
\infty$ we have
$$
\int^{+\infty}_0{e^{-hs}\over (\rho^2+s^2)^{d-1\over 2}}\simeq {1\over h~\rho^{d-1}}-{(d-1)\over h^3}{1\over \rho^{d+1}}~~,
$$
$$
\int^{+\infty}_0e^{-hs}\ln (\rho^2+s^2)ds\simeq {1\over h}\ln \rho^2+{2\over
h^3\rho^2}
$$
and the correlators (\ref{2.3}) behave as
\begin{equation}
K_{(h)}\simeq - {1\over h^2} K_D~~,~~K_D^{d\geq 1}={1\over \Sigma_d}{1\over \rho^{d+1}}~~.
\label{2.4}
\end{equation}
where $K_D=-{1\over 2}\partial_z\partial_{z'} G_D |_{z=z'=0}$ is the 
correlator arising on the boundary in the Dirichlet  problem
with Green's function $G_D$ (\ref{2.1}).
So in the limit of large $h$ the boundary operator ${\cal O}^{(h)}$ behaves as
${\cal O}^{(h)}\simeq {1\over h}{\cal O}_{D}$, where ${\cal O}_{D}$
is the operator arising on the boundary in the Dirichlet problem.
Note that the Dirichlet correlator (\ref{2.4}) induced on $d$-dimensional
boundary is related to the Neumann correlator induced on $(d+2)$-dimensional
boundary: $K^d_D=-2\pi K^{d+2}_N$. This indicates the duality
of the strong and weak coupling regimes: 
the Neumann problem on $(d+3)$-dimensional
space is dual to the Dirichlet problem on $(d+1)$ dimensions.

The operators ${\cal O}^{(h)}(P)$ have dimension ${(d-1)\over 2}$. 
Therefore, their correlators are likely to behave as ${1\over \rho^{d-1}}$ for
small $\rho$. However, for a fixed coupling $h$ in the limit $\rho\rightarrow 0$ the correlators (\ref{2.3}) possess a series of the divergent terms
\begin{equation}
K^d_{(h)}\simeq {1\over (d-1)}{1\over \Sigma_d}\left(
{1\over \rho^{d-1}}+{b_0^{(d)}h\over \rho^{d-2}}+...+{b_k^{(d)}
h^k\over \rho^{d-k-1}}+...+b^{(d)}_{d-1}h^{d-1} \ln (\rho h) \right)~~,
\label{S}
\end{equation}
where $\{ b_k^{(d)}\}$ are some coefficients.
For $(d-1)$ even only even powers of $\rho$ can appear in the series (\ref{S})
so in this case the coefficients $b_{2k}^{(d)}$ vanish.
In a few particular cases the series (\ref{S}) reads
\begin{eqnarray}
&&K^1_{(h)} \simeq -{1\over 4\pi}\ln \rho^2~~, \nonumber \\
&&K^2_{(h)} \simeq {1\over \Sigma_2}\left( {1\over \rho}+h\ln (\rho h)\right)
\nonumber \\
&&K^3_{(h)} \simeq {1\over 2\Sigma_3}\left( {1\over \rho^2}-h^2\ln (\rho h)\right)
\nonumber \\
&&K^4_{(h)} \simeq {1\over 3\Sigma_4}\left( {1\over \rho^3}-{h\over \rho^2}+{h^3\over 2}\ln (\rho h)\right)
\nonumber \\
&&K^5_{(h)} \simeq {1\over 4\Sigma_5}\left( {1\over \rho^4}-{h\over 2\rho^2}+{h^4\over 6}\ln (\rho h)\right)
\nonumber \\
&&K^6_{(h)} \simeq {1\over 5\Sigma_6}\left( {1\over \rho^5}-{2h\over 3\rho^4}-
{h^3\over 6\rho^2}+{h^4\over 9\rho}+{h^5\over 24}
\ln (\rho h)\right)
\label{S1}
\end{eqnarray}
The expansion (\ref{S}), (\ref{S1}) is valid for $(\rho h)<<1$. The leading
contribution is given by the Neumann correlator (\ref{2.4}) arising
in the free regime. In the opposite case, when $(\rho h)>>1$, the correlators
(\ref{2.3}) tend to the strong coupling expression (\ref{2.4}).
We see that in the region $(\rho h)<<1$ the weak coupling regime is restored. In this regime an operators ${\cal O}^{(h)}(P)$ can be viewed
as a (perturbative in $h$) composite
of operators of dimensions ${(d-k-1)\over 2}$. The contribution of each such 
operator to the correlator (\ref{S}), (\ref{S1}) comes with weight $h^k$.
In the region $(\rho h)>>1$ we have a strong coupling (Dirichlet)
phase. The parameter $h$ characterizes the size of the region separating
the free (Neumann) and the strong coupling (Dirichlet) phases.

The correlators (\ref{2.3}) being considered as 
functions of the geodesical distance $\rho$ possess  interesting
recurrent relations. Consider a set of operators $\Delta_n=\rho^{-(n-1)}
\partial_\rho(\rho^{n-1}\partial_\rho )$. $\Delta_n$ is radial part
of the Laplace operator in $d$ dimensions. A combination of these
operators is a first-order differential operator: 
$\Delta_n-\Delta_{n-k}={k\over \rho}\partial_\rho$.
It is easy to see using the property $\Sigma_{d+2}={2\pi\over d+1}\Sigma_d$
that
\begin{equation}
\rho^{-1}\partial_\rho K^d_{(h)}=-2\pi K^{d+2}_{(h)}~~, ~d\geq 1~~.
\label{2.5}
\end{equation}
This relation is independent of $h$. In the limit of strong coupling
we find that $K^d_{(h)}\simeq {2\pi\over h^2}K^{d+2}_{h=0}=-{1\over h^2}K_D^{d}$ and we have the relations
$$
\rho^{-1}\partial_\rho K^d_N=K^{d+2}_D~~,~~K_D^d=-2\pi K^{d+2}_N
$$
which establish the duality between the Neumann and Dirichlet problems
on flat space.

The correlators $K^d_{(h)}(\rho )$ are defined on $d$-dimensional
space $R^d$. However, one can consider them on space of arbitrary dimension.
More precisely, on a space ${\cal C}=R^n$  one can define
all tower
of correlators $\{ K^d_{(h)}(\rho ),~d\geq 1\}$ where $\rho$ is the distance measured on ${\cal C}=R^n$. It can be done in the following way. Consider two
points $P$ and $P'$ on ${\cal C}=R^n$ and a space ${\cal M}=R^{d+1}$ which boundary
${\cal B}=\partial {\cal M}$ intersects the space ${\cal C}=R^n$ 
in such a way 
that $P$ and $P'$ are lying on $\cal B$. Then a field theory on $\cal M$
induces correlator $K^d_{(h)}$ for any pair of points
on $\cal B$ and particularly for
the points $P$ and $P'$. Using now the group of symmetry of the 
space ${\cal C}$
and moving $\cal M$ along  ${\cal C}$ one can define the correlator
$K^d_{(h)}(\rho )$ for any pair of points on $\cal C$.
This offers an interesting interpretation for the correlation function 
(\ref{S}). It arises for a pair of points of the space ${\cal B}=R^d$
lying on intersection of $\cal B$ with boundaries of the set of planes
$\{{\cal P}^k=R^{d-k+1},~k=0,...,d-1 \}$
with the Neumann boundary problem considered 
on each plane. Then each plane ${\cal P}^k$ induces
$N$-correlator $\sim {1\over \rho^{d-k-1}}$ for this pair of points,
and the total correlator (\ref{S}) is just a sum over all these contributions.

We finish this Section with a brief comment.  Consider the following integral
$$
\int d^dx\int d^dy f(x){1\over |x-y|^n}f(y)~~.
$$
It is well-defined for $n<d$ and is divergent when $x\rightarrow y$
for $n\geq d$. The integrals of this type appear
in our boundary action. It follows that the boundary action
$W^{(h)}=\int\limits_{~~\cal B}\!\!\int g(P)K_{(h)}(P,P')g(P')$
is finite for any finite $h$ since the leading divergence in the correlator
$K_{(h)}$ is $1\over \rho^{d-1}$. On the other hand, the correlation function
$K_D$ in the Dirichlet phase  behaves as $1\over \rho^{d+1}$ and the action
$W^{D}=\int\limits_{~~\cal B}\!\!\int f(P)K_{D}(P,P')f(P')$
should be regularized. The simplest way to do this
is to consider the action
\begin{equation}
W_{reg}^D=\int\limits_{~~\cal B}\!\!\int f(P)K_{D}(P,P')f(P')
-\int\limits_{~~\cal B}\!\!\int f(P)K_{D}(P,P')f(P)
\label{R}
\end{equation}
which is obviously finite. The last term in (\ref{R}) can be re-written
as follows
$$
\tilde{h}\int_{\cal B}f^2(P)~,~~\tilde{h}=-c\int^{+\infty}_0r^{-2}dr~~,
$$
where $c$ is a (positive) constant. We see that the counter-term to be added
to $W^D$ takes the form of local boundary term as in (\ref{18}).
Alternatively, one may use analyticity method \cite{17} to regularize
the Dirichlet action. It involves the identity $\int_0^{+\infty}r^\lambda dr=0$
which is proved by analyticity in $\lambda$. A regularization of the correlation functions has been also discussed in \cite{6}, \cite{7}.

\bigskip

\section{Flat space with spherical boundary}
\setcounter{equation}0
In this Section we study the boundary correlation functions 
described in Section 2 when
manifold $\cal M$ is a domain of $(d+1)$-dimensional flat space $R^{d+1}$
with boundary $\cal B$ being $d$-dimensional sphere $S^d$ of radius $R$.
We start with analysis of the simplest  $d=1$ case.

\medskip

{\bf 4.1 Dirichlet problem on $R^2$ with boundary $S^1$} \\
Let $P$ and $M$ be points on $R^2$ with polar coordinates $(\rho_0,\phi ')$
and $(\rho , \phi )$ respectively. We consider the domain
inside the circle $S^1$ of radius $R$, i.e. $\rho_0,\rho \leq R$.
Green's function takes the form
\begin{equation}
G(P,M)={1\over 2\pi}\ln {1\over r(P,M)}+v(P,M)~~,
\label{3.1}
\end{equation}
where $r(P,M)=\sqrt{\rho^2+\rho^2_0-2\rho\rho_0\cos (\phi-\phi ')}$
is distance between the points $M$ and $P$ and $v(P,M)$ is harmonic function which is regular everywhere in the domain.  $v(P.M)$ 
is determined by the Dirichlet condition $G(P,M)=0$ if $P$ lies on $\cal B$. In order to construct
$v(P,M)$ consider point $M^*$ with coordinates $(\rho^*={R^2\over \rho}, \phi )$. It is conjugate  to the point $M$. The distance $r_1$ between  
$P$ and $M^*$ satisfies the relation
\begin{equation}
r^2_1={R^2\over \rho^2}\left(r^2+\rho^2({\rho^2_0\over R^2}-1)+R^2-\rho^2 
\right)~~.
\label{r1}
\end{equation}
If $P$ lies on $S^1$ ($\rho_0=R$) then we have $r_1={Rr\over \rho}$. This implies that the Dirichlet Green's function $G_D$ takes the well-known form
\begin{equation}
G_D={1\over 2\pi}\left( \ln {1\over r}-\ln {R\over \rho r_1}\right)~~.
\label{3.3}
\end{equation}
It is easy to find that
\begin{equation}
\left( \partial_n r\right)_{\rho_0=R}=
{R^2+r^2-\rho^2\over 2R r}~~,~~\left( \partial_n r_1\right)_{\rho_0=R}=
{\rho^2+r^2-R^2\over 2\rho r}
\label{3.4}
\end{equation}
and the normal derivative of $G_D$ is
$$
\partial_n G_D|_{\rho_0=R}=-{1\over 2\pi R}{(R^2-\rho^2)\over r^2}~~.
$$
Thus, the solution of the Dirichlet problem $u|_{\rho_0=R}=f(\phi )$
 takes the form (see eq.(\ref{10}))
\begin{equation}
u_D(\rho, \phi )={1\over 2\pi R}\int_0^{2\pi}{R^2-\rho^2\over R^2+\rho^2-2R\rho \cos (\phi-\phi ')} f(\phi ') Rd\phi '
~~.
\label{3.5}
\end{equation}
This formula is known as Poisson's integral for a circle \cite{14}.

Calculating the second normal derivative of $G_D$ on the boundary we obtain according to eq.(\ref{16}) the correlation function on the boundary
${\cal B}=S^1$
\begin{equation}
<{\cal O}_D(\phi ){\cal O}_D(\phi ' )>= K_D(\phi , \phi ')=-{1\over 8\pi R^2\sin^2 ({\phi -\phi '\over 2})}~~.
\label{3.6}
\end{equation}
This is in agreement with 
that the boundary operator ${\cal O}_D$ has dimension $1$.

\bigskip

{\bf 4.2 Dirichlet problem on $R^{d+1}$ with boundary $S^d$} \\
The above two-dimensional result can be easily generalized for the Dirichlet
problem on a
domain of $(d+1)$-dimensional space $R^{d+1}$ with spherical boundary $S^d$.
The metric on $R^{d+1}$ takes the form
$$
ds^2=d\rho^2+\rho^2 d\omega^2~~,
$$
where $d\omega^2$ is the metric on the $d$-dimensional unit sphere.
In the spherical coordinate system with origin in the 
center of the sphere consider two points $P$ and $M$ with coordinates
$(\rho_0, \theta_i')$ and $(\rho , \theta_i)$ respectively, $\theta_i~,i=1, ...
d$ are angle coordinates  on the sphere $S^d$.
The distance $r$ between them
is found from the relation
$$
r^2=\rho^2_0+\rho^2-2\rho\rho_0\cos \gamma~~,
$$
where $\gamma$ ($0\leq \gamma\leq\pi$) is the geodesical distance measured on the $d$-dimensional unit sphere.

Green's function satisfying the Dirichlet boundary condition
on $S^d$ takes the form
\begin{equation}
G_D={1\over (d-1)}{1\over \Sigma_d}\left({1\over r^{d-1}}-({R\over \rho })^{d-1}
{1\over r_1^{d-1}}\right)~~,
\label{3.8}
\end{equation}
where $r_1$ is the distance between the points $P$ and 
$M^*$ defined in the same way as in the previous subsection.
Note that $r$ and $r_1$ satisfy the same relations (\ref{r1}) and (\ref{3.4})
as in two-dimensional case. Proceeding in the same way as in  Section 4.1
we find a generalization
of Poisson's integral (\ref{3.5}) in higher dimensions
\begin{equation}
u_D(\rho , \theta_i)={1\over \Sigma_d}{1\over R}\int_{S^d}
{(R^2-\rho^2)\over (R^2+\rho^2-2R\rho \cos\gamma )^{d+1\over 2}}
f(\theta ')R^d d\mu (\theta )~~,
\label{3.9}
\end{equation}
where $R^dd\mu (\theta )$ is measure on $S^d$, $\gamma$ is the geodesical distance on the unit sphere between points with coordinates $\theta$ and $\theta '$.
The correlation function defined on the boundary $S^d$ then reads
\begin{equation}
K_D(\theta ,\theta ')=-{1\over \Sigma_d}{1\over (4R^2\sin^2{\gamma\over 2})^{d+1\over 2}}~~.
\label{3.10}
\end{equation}
It is a higher-dimensional 
generalization of the expression (\ref{3.6}).

\bigskip

{\bf 4.3 Massive case and the limit $R\rightarrow 0$} \\
If the field $u$ is massive then we have the equation
\begin{equation}
(\Box -m^2)u=0
\label{3.11}
\end{equation}
instead of the Laplace equation (\ref{2}).  In $(d+1)$-dimensional
flat space ${\cal M}=R^{d+1}$ the fundamental solution of this equation is found
to take the form
\begin{equation}
g_d(m,r)={1\over (2\pi )^{d+1\over 2}}({m\over r})^{d-1\over 2}K_{d-1\over 2}
(mr)~~,
\label{3.12}
\end{equation}
where $r$ is the distance between points $P$ and $M$ in $R^{d+1}$.
For even $d$ the expression (\ref{3.12}) is particularly simple 
taking into account that the modified Bessel function $K_\nu$ is
$$
K_{n+{1\over 2}}(z)=\sqrt{\pi\over 2z}e^{-z}\sum_{k=0}^n{(n+k)!
\over k! (n-k)!(2z)^k} ~~
$$
for integer $n$.

The expression (\ref{3.12}) can be obtained as a result of applying
the formula
$$
g_d(r)=\int_0^{+\infty}ds {\cal K}_{R^{d+1}}(s,r)~~,
$$
where ${\cal K}_{R^{d+1}}(s,r)$ is the heat kernel function on $R^{d+1}$
defined as solution of the equation
$$
\partial_s{\cal K}=(\Box-m^2){\cal K}~~.
$$
In $R^{d+1}$ the heat kernel is known to take the form
\begin{equation}
{\cal K}_{R^{d+1}}(s,r)={1\over (4\pi s)^{d+1\over 2}}e^{-m^2s}e^{-{r^2\over 4s}}~~.
\label{3.13}
\end{equation}

Green's function for the operator (\ref{3.11}) with the Dirichlet condition on sphere can not be found using the method of images so useful in the massless
case. It is, however, possible to construct an auxiliary Green's
function $\tilde{G}_D(P,P')$ which vanishes when both
points $P$ and $P'$ lie on the boundary
\begin{equation}
\tilde{G}_D(m,r)=g_d(m,r)-g_d({R\over\rho}m,{\rho \over R}r_1)~~,
\label{3.14}
\end{equation}
$\rho$ and $r_1$ are
defined as in eq.(\ref{3.8}).
Near the boundary the true Green's function $G_D(P,P')$ \\
$\simeq $
$\tilde{G}_D(P,P')+$$
\alpha_0(\gamma )+\alpha_1(\gamma )(R-\rho )$$+ \alpha_1'(\gamma )(R-\rho ' )$.
Therefore, the knowledge of $\tilde{G}_D(P,P')$ is enough for our goals.

Using the eq.(\ref{3.4}) and the identity $x\partial_x K_\nu(x)-\nu K_nu 
(x)=-xK_{\nu+1}
(x)$ for the modified Bessel functions and
 calculating the normal
derivatives of the expression (\ref{3.14}) we find the boundary correlation function
\begin{eqnarray}
&&K_D^d(m,\gamma )=-{1\over (2\pi )^{d+1\over 2}}({m\over 2R\sin {\gamma \over 2}}
)^{d+1\over 2}K_{d+1\over 2}(2mR\sin {\gamma\over 2})\nonumber \\
&&+{1\over (2\pi )^{d+1\over 2}}{m^4\over 4R^2}({m\over 2R\sin{\gamma\over
2}})^{d-5\over 2}K_{d-1\over 2}(2mR\sin{\gamma\over 2})~~
\label{3.15}
\end{eqnarray}
corresponding to the massive theory (\ref{3.11}) in the bulk.
The first term in (\ref{3.15}) is the leading one for small $\gamma$.
It takes the same form as for plane boundary. The second term in (\ref{3.15})
collects the effects of curvature of the boundary. For small $\gamma$
it is proportional to ${m^2\over R^2}(R\sin{\gamma\over 2})^{3-d}$
and leads to new divergent terms only for $d>3$.

It is of interest to consider the limit $R\rightarrow 0$ in the
expression (\ref{3.15}). Then the Bessel function  can be approximated
by its value at small values of argument. The leading term 
is given by the  correlator (\ref{3.10})
of the massless case. This is not surprising because the eq.(\ref{3.11}) is
effectively massless for small values of the radial coordinate. However,
there also appears the whole series of the  subleading divergent
 terms 
when we take the limit of small radius $R$ in (\ref{3.15}) . These terms 
are proportional to powers of mass
$m$.  In order to accomplish the limit $R\rightarrow 0$
we introduce an operator $\tilde{\cal O}_D(P)$ having the same dimension 
$d+1\over 2$ as the operator ${\cal O}_D(P)$.
For small $R$ 
we have ${\cal O}_D(P)\simeq ({R\over a})^{-{d+1\over 2}}\tilde{\cal O}_D(P)$
where $a$ is some (finite) scale.
The limit $R\rightarrow 0$ for the correlation function of the 
operators $\tilde{\cal O}_D(P)$ is well defined 
\begin{equation}
<\tilde{\cal O}_D(P)\tilde{\cal O}_D(P')>_{R\rightarrow 0}= -{1\over \Sigma_d}
{1\over (2a\sin{\gamma\over 2})^{d+1}}
\label{3.15'}
\end{equation}
and coincides with the  the correlation function (\ref{3.10}) 
on sphere $S_d$ of 
radius $a$ in the massless case.

\bigskip

{\bf 4.4 Dirichlet problem on $R^{d+1}\times \Sigma$} \\
The results of Section 4.3 are useful when we consider the Dirichlet problem 
on a domain of  a product space $R^{d+1}\times \Sigma$
with boundary ${\cal B}=S^d\times \Sigma$, where $\Sigma$ is compact  
manifold with coordinates $\{\chi\}$.
Let the functions $\{Y_n(\chi )\}$ form the orthonormal basis of eigenfunctions of the Laplace operator $\Box_\Sigma$ considered on the compact space $\Sigma$, i.e. we have
\begin{eqnarray}
&&\Box_\Sigma Y_n=-\lambda_n^2Y_n \nonumber \\
&&\int_\Sigma Y_n(\chi )Y_m(\chi )d\mu (\chi )=\delta_{n,m}~~.
\label{Y}
\end{eqnarray}
The field function $u$ being considered on $R^{d+1}\times \Sigma$
expands with respect to this basis
$$
u(P)=\sum_n Y_n(\chi )u_n~~,
$$
where $\{u_n\}$ are functions on $R^{d+1}$ satisfying the equation 
(\ref{3.11}) with ``mass'' $m=\lambda_n$. The Dirichlet boundary condition 
consists in fixing the infinite set of functions $\{f_n\}$:
$$
u|_{\cal B}=f(\chi ,\theta )=\sum_n Y_n(\chi )f_n(\theta )~~,
$$
where $\{\theta \}$ are coordinates on $S^d$.

Green's function expands with respect to the basis $\{Y_n\}$ as follows
\begin{equation}
G(P,P')=\sum_n Y_n(\chi )Y_n(\chi ')G_n
\label{3.16}
\end{equation}
where $G_n$ is Green's function for the operator (\ref{3.10})
on $R^{d+1}$ with mass $m=\lambda_n$.

Considering on the boundary the source term
$$
\int_{\cal B}f(\chi ,\theta ){\cal O}_D(\chi , \theta )
d\mu (\chi )d\mu (\theta )
$$
we obtain the correlation function for the boundary operator
${\cal O}_D(\chi , \theta )$
\begin{eqnarray}
&&<{\cal O}_D(\chi , \theta ){\cal O}_D(\chi ' , \theta ' )>=
K_{S^d\times \Sigma} (\chi , \theta , \chi ' , \theta ' ) \nonumber \\
&&=\sum_n Y_n(\chi )Y_n(\chi ')K_D(\lambda_n, \gamma )~~,
\label{3.17}
\end{eqnarray}
where $K_D(\lambda ,\gamma)$ is the correlator (\ref{3.15}).
In general it is rather complicated function. We are, however, interested
in considering the limit of (\ref{3.17}) when radius
$R$ of the sphere $S^d$ goes to zero. In this limit the function $K_D(\lambda_n, \gamma )$ becomes independent of $\lambda_n$.
Introducing new operators $\tilde{\cal O}_D(\chi , \theta )$
such that ${\cal O}_D(P)\simeq ({R\over a})^{-{d+1\over 2}}\tilde{\cal O}_D
(P)$ for small $R$,  we obtain the following 
remarkable factorization 
\begin{equation}
\tilde{K}_{S^d\times\Sigma }(\chi , \theta , \chi ' , \theta ' )=
\delta_\Sigma (\chi-\chi ')K_{S^d}(\theta , \theta ')~~,
\label{3.18}
\end{equation}
where $K_{S^d}(\theta , \theta ')$ is the correlator (\ref{3.10}), 
(\ref{3.15'})
arising on sphere $S^d$ of radius $a$. The correlator (\ref{3.18})
describes motion of a quantum particle which does not
propagate along the component $\Sigma$ of the space $S^d\times \Sigma$.

\bigskip

{\bf 4.5 Neumann problem on $R^2$} \\
Solving the elliptic problem with the Neumann condition imposed
on spherical boundary 
we need to modify the definition of Green's function. More precisely,
Green's function $G_N(P,P')$ of the Neumann problem 
should be a solution of the equation
\begin{equation}
\Box G_N(P,P')=C
\label{3.19}
\end{equation}
subject to the Neumann boundary condition
\begin{equation}
\partial_n G_N|_{P\in {\cal B}}=0~~.
\label{3.20}
\end{equation}
The right hand side of the equation (\ref{3.19}) is a constant $C$. Therefore,
we still can apply Green's formula (\ref{6}) and obtain the expression
(\ref{10}) for the solution $u_N(P)$ of the Neumann problem
taking into account that $u_N(P)$ is determined up to an irrelevant constant.
In fact, the constant $C$ in (\ref{3.19}) is not arbitrary and is determined 
 by the requirement of consistency of the boundary condition
(\ref{3.20}). In two dimensions we find that $C={1\over \pi R^2}$ and Green's function takes the form
\begin{equation}
G_N(P,P')={\rho_0^2\over 4\pi R^2}+{1\over 2\pi}\left(\ln{1\over r}
+\ln {R\over \rho r_1}\right)~~,
\label{3.21}
\end{equation}
where $(\rho_0,\phi '),~(\rho , \phi )$ are coordinates of the points $P$ and $P'$ respectively. Definitions of $r$ and $r_1$ are the same as in Section 4.1.
By means of  simple calculations using eq.(\ref{3.4}) one shows 
that the condition
(\ref{3.20})  indeed fulfills for Green's function (\ref{3.21}).
Considering (\ref{3.21}) when both points lie 
on the boundary $\cal B$ ($\rho_0=\rho=R$)
we obtain according to (\ref{16}) the Neumann correlation function
on ${\cal B}=S^1$
\begin{equation}
K_N(\phi , \phi ')=-{1\over 4\pi }\ln \sin^2{\phi-\phi '\over 2}~~,
\label{3.22}
\end{equation}
where we neglected all irrelevant constants.

\bigskip

{\bf 4.6 Neumann problem on $R^{d+1},~d\geq 2$} \\
In higher dimensions we search the solution of the equation (\ref{3.19})
in the form \cite{15}
\begin{equation}
G_N(P,P')=\alpha^{-1} \rho^2_0+{1\over (d-1)}{1\over \Sigma_d}{1\over r^{d-1}}
+v(P,P')~~,
\label{3.23}
\end{equation}
where $v(P,P')$ is harmonic function, $\Box v=0$, which is regular everywhere
in the domain. Note that in $R^{d+1}$ the function $\rho^2_0$ is a solution of the equation $\Box \rho^2_0=2(d+1)$. The condition (\ref{3.20})
is consistent if we put
$\alpha=2(d+1)R^{d+1}\Sigma_{d+1}$ in (\ref{3.23}) and $C=2(d+1)\alpha^{-1}$
in (\ref{3.19}).

In 3-dimensional case Green's function with the Neumann condition on
$S^2$ takes the following form \cite{15}
\begin{equation}
G_{N}={\rho_0^2\over 8\pi R^3}+{1\over 4\pi r}+{1\over 4\pi r_1}{R\over \rho}+w~~,
\label{3.24}
\end{equation}
where
\begin{equation}
w=-{1\over 4\pi R}\ln \left(r_1+{R^2\over \rho}-\rho_0\cos \gamma \right)
\label{3.25}
\end{equation}
is harmonic function, $\Box w=0$. The calculation of 
normal derivatives 
$$
\partial_n \rho^2_0|_{\rho_0=R}=2R~,~~\partial_n w|_{\rho_0=R}=-{1\over 4\pi R}
({1\over R}-{1\over r})
$$
$$\partial_n ({1\over 4\pi r}+{1\over 4\pi r_1}{R\over \rho})|_{\rho_0=R}=
-{1\over 4\pi Rr}
$$
shows that
the condition (\ref{3.20}) fulfills
for Green's function (\ref{3.24}).

The boundary correlation function corresponding, in accord with (\ref{16}),
to Green's function (\ref{3.24}) reads
\begin{eqnarray}
&&K^{d=2}_N={1\over 2}G_N(P,P')|_{P,P'\in {\cal B}} \nonumber \\
&&={1\over 8\pi R\sin{\gamma\over 2}}-{1\over 8\pi R}\ln
\sin^2{\gamma\over 2}~~,
\label{3.26}
\end{eqnarray}
where $\gamma$ ($0\leq \gamma \leq \pi$) is geodesical distance between 
$P$ and $P'$ on the unit two-dimensional sphere. 
If the spherical coordinates of the 
points $P$ and $P'$ are $(\phi ,\theta )$ and $(\phi ', \theta ')$ respectively
then $\gamma$ is found from the relation
\begin{equation}
\sin^2{\gamma\over 2}=\sin^2({\theta-\theta '\over 2})+\sin\theta\sin\theta '
\sin^2({\phi-\phi '\over 2})~~.
\label{gamma}
\end{equation}

The first term in (\ref{3.26}) is what we could anticipate basing on the analysis of the Neumann problem on plane boundary (see eq.(\ref{N})).
Surprisingly, we obtain also the logarithmic term in the correlator
(\ref{3.26}). 
It arises due to curvature of the boundary and
could not be anticipated
just from the analysis of the  plane boundary problem.

The generalization of the result (\ref{3.26}) to higher
dimensions, $d>2$, is 
technically difficult. As a conjecture we, however,
propose that the general structure of the Neumann correlation function for
$d\geq 3$ includes the series
\begin{equation}
K^d_N={1\over (d-1)}{1\over \Sigma_d}\left(
{1\over \sigma^{d-1}}+{c_1^{(d)}R^{-1}\over \sigma^{d-2}}+...+
{c_k^{(d)}R^{-k}\over \sigma^{d-k-1}}+...+c_{d-1}^{(d)}R^{1-d}
\ln \sigma\right)
\label{3.27}
\end{equation}
with respect to variable $\sigma=2R\sin {\gamma\over 2}$,
$\gamma$ is geodesical distance measured on unit sphere
$S^d$ and $c^{(d)}_k$ are constant
coefficients. Note that all terms in the series are proportional to 
$R^{(1-d)}$. The expression (\ref{3.27}) is worth comparing with
the Eq.(\ref{S}). It seems that the behavior (\ref{3.27}), (\ref{S})
is typical for the correlation function in the free (Neumann) phase.
In both cases the expected term $\sim {1\over \sigma^{d-1}}$
gets modified by a series with respect to dimensional parameter
(additional to the geodesical distance $\sigma$) which is $h$
in Eq.(\ref{S}) and $R$ in (\ref{3.27}).

In the massive case the arguments similar to that of Section 4.3 say that
taking the limit $R\rightarrow 0$ one gets the massless Neumann
correlators (\ref{3.26}), (\ref{3.27}) provided that we rescaled the Neumann 
boundary operators: ${\cal O}_N=({R\over a})^{-{d-1\over 2}}\tilde{\cal O}_N$.
As a consequence of this, the correlator arising in the Neumann problem
on a product space $R^{d+1}\times \Sigma$ ($\Sigma$ is a compact manifold)
possesses in the limit
of small $R$ ($R$ is radius of the boundary in $R^{d+1}$) the factorization
\begin{equation}
\tilde{K}_N[S^d\times\Sigma ]=\delta_\Sigma ~K_N[S^d]
\label{3.28}
\end{equation}
similar to what we had for the Dirichlet case (\ref{3.18}).

\bigskip

\section{Anti-de Sitter (AdS) space with boundary \\
at infinity}
\setcounter{equation}0

\bigskip

{\bf 5.1 Dirichlet problem on $AdS_{d+1}$} \\
General analysis of AdS spaces we start with the consideration of
two-dimensional AdS space
$H^2$. It is space of constant negative curvature $-{2\over l^2}$ described by the metric
\begin{equation}
ds^2=l^2(dx^2+\sinh^2xd\phi^2)~~,
\label{4.1}
\end{equation}
where the coordinates $(x, \phi )$ run in the limits
$0\leq \phi \leq 2\pi$, $0\leq x\leq +\infty$. Boundary $\cal B$ of the space
$H^2$ is a circle $S^1$ lying at the infinite value of the coordinate
$x$. Considering the Laplace equation on $H^2$ we
impose the Dirichlet boundary condition $u|_{x\rightarrow +\infty}=f(\phi)$
on  the field function $u$ at infinity.

Green's function which satisfies the Dirichlet condition 
$G_{H^2}(x,x',\phi ,\phi ')=0$ at infinity $(x\rightarrow \infty )$ takes the following form
\begin{equation}
G_{H^2}=-{1\over 2\pi}\ln \tanh {\sigma \over 2l}~~,
\label{4.2}
\end{equation}
where $\sigma$ is geodesical distance between points $(x,\phi )$ and $(x', \phi ')$ on $H^2$, it can be found from the relation
\begin{equation}
\cosh {\sigma\over l}=\cosh x \cosh x'-\sinh x \sinh x' \cos (\phi-\phi ')~~,
\label{4.3}
\end{equation}
which is a hyperbolic analog of the relation (\ref{gamma}) for the
geodesical distance on $2$-sphere.

Fixing the point $(x', \phi ')$ consider the limit when the second point $(x, \phi )$ goes to infinity ($x\rightarrow +\infty $). In this limit we have 
$G_{H^2}={1\over \pi}e^{-{\sigma\over l}}$. On the other hand, we find from the
relation (\ref{4.3}) that
\begin{equation}
e^{\sigma \over l}\simeq e^x \Delta (x', \phi-\phi ')~,
~~\Delta (x', \phi-\phi ')=\cosh (x')-\sinh x' \cos (\phi-\phi ')~~.
\label{4.4}
\end{equation}
Hence Green's function for $x \rightarrow +\infty$ reads
\begin{equation}
G_{H^2}={1\over \pi}{e^{-x}\over \Delta}~~.
\label{4.5}
\end{equation}
Normal derivative $\partial_n=l^{-1}\partial_x$ of this expression
\begin{equation}
\partial_nG_{H^2}=-{1\over \pi l}{e^{-x}\over \Delta}~~
\label{4.6}
\end{equation}
is exponentially decreasing for $x$ going to infinity.

The measure $d\mu=l\sinh x d\phi$ arising on the boundary $\cal B$ contains the exponentially growing factor.
Therefore, applying the general formula (\ref{10}) we find the
solution of the elliptic problem on $H^2$ with Dirichlet boundary condition at 
infinity in the form
\begin{equation}
u_D(x',\phi ')={1\over 2\pi}\int^{2\pi}_0{f(\phi )d\phi
\over \cosh (x')-\sinh x' \cos (\phi-\phi ')}
\label{4.7}
\end{equation}
that is a hyperbolic version of Poisson's integral (\ref{3.5}).
Note, that (\ref{4.7}) gives us the solution of the Laplace equation
on the whole space $H^2$.

Putting $x'=0$ in (\ref{4.7}) we find that $u(0,\phi ')={1\over 2\pi}
\int^{2\pi}_0f(\phi )d\phi$ for value of the harmonic function $u$
in the center of the space $H^2$. It is exactly the same
value which we get in the case of flat disk (putting $\rho_0=0$
in (\ref{3.5})). On the other hand, considering $x'\rightarrow +\infty$
in (\ref{4.7})
we find that the kernel in the integral (\ref{4.7})
reproduces $\delta$-function $\delta (\phi-\phi ')$ what is in
agreement with the imposed boundary condition.

Taking the limit $x'\rightarrow +\infty$ in (\ref{4.6}) and calculating
the normal derivative $\partial_{n'}=l^{-1}\partial_{x'}$ we find
that
$$
\partial_{n'}\partial_n G_{H^2}={1\over \pi l^2}{e^{-(x+x')}\over \sin^2({\phi-
\phi'\over 2})}
$$
and the action $W[u_D]$ (\ref{13})
reads
$$
W[u_D]=-{1\over 8\pi}\int_0^{2\pi}\int_0^{2\pi}f(\phi ){1\over \sin^2({\phi-
\phi'\over 2})}f(\phi ')d\phi d\phi '~~.
$$
This results in the correlation function
\begin{equation}
K_{D}(\phi ,\phi ')=-{1\over 8\pi a^2}{1\over \sin^2({\phi-
\phi'\over 2})}
\label{4.8}
\end{equation}
arising on the boundary ${\cal B}=S^1$ with measure $ad\phi$. 
It exactly reproduces the correlation function (\ref{3.6}) appearing in the Dirichlet problem on flat disk $D^2$.
It is as expected since spaces $H^2$ and $D^2$ are conformally related.

These results can be generalized for higher-dimensional AdS spaces.
The $(d+1)$-dimensional hyperbolic space $H^{d+1}$ is described by the metric
\begin{equation}
ds^2=l^2(dx^2+\sinh^2xd\omega^2)~~,
\label{4.9}
\end{equation}
where $d\omega^2$ is metric of $d$-dimensional unit sphere $S^d$,
set of angles $\{\theta_i,~i=1,...,d\}$ being coordinates on $S^d$.
The geodesical distance $\sigma$  between two points $(x,\theta )$ and
$(x', \theta ')$ on $H^{d+1}$ can be found from the relation
\begin{equation}
\cosh {\sigma\over l}=\cosh x \cosh x'-\sinh x \sinh x' \cos \gamma
\label{4.10}
\end{equation}
analogous to eq.(\ref{4.3}), $\gamma$ is geodesical distance measured on 
unit sphere
$S^d$. Solution of the Laplace equation $\Box u=0$ on $H^{d+1}$
for large values of $x$ behaves as follows \cite{3}
\begin{equation}
u(x, \theta )\simeq u_0(\theta )+u_1 (\theta )e^{-dx}~~,
\label{4.11}
\end{equation}
where $u_0$ and $u_1$ are some functions on $S^d$. Demanding 
that $u(x, \theta )$ is regular at $x=0$ one finds that only one    of these functions is independent.

The
Dirichlet boundary condition at infinity reads 
$$
u(x,\theta )|_{x\rightarrow \infty}=f(\theta )~~.
$$
As in two-dimensional case we  fix point $(x' , \theta ')$
and consider the other point $(x, \theta )$ approaching
infinity. Green's function in $H^{d+1}$ space vanishing at infinity
has the following integral representation
\begin{equation}
G_{H^{d+1}}={l^{1-d}\over \Sigma_d}\int_{\sigma\over l}^{+\infty}{dx\over \sinh^dx}~~.
\label{G}
\end{equation}
We, however, need to know only behavior of Green's function for large values of $\sigma$
\begin{equation}
G_{H^{d+1}}\simeq {2^d\over d\Sigma_d}{1\over l^{d-1}}e^{-d{\sigma\over l}}~~.
\label{4.12}
\end{equation}
We find from (\ref{4.10}) the expression
\begin{equation}
e^{\sigma \over l}\simeq e^x \Delta (x', \theta ,\theta ')~,
~~\Delta (x', \theta ,\theta ')=\cosh x'-\sinh x' \cos \gamma ~~,
\label{4.13}
\end{equation}
valid for large 
value of $x$,
$\gamma$ is geodesical distance on $S^d$ between points with coordinates
$\theta$ and $\theta '$. Equation (\ref{4.13}) is a higher-dimensional
generalization of eq.(\ref{4.4}). Combining (\ref{4.12}) and (\ref{4.13}) we get Green's function
for large values of $x$
\begin{equation}
G_{H^{d+1}}\simeq {2^d\over d\Sigma_d}{1\over l^{d-1}}\Delta^{-d}
e^{-dx}~~.
\label{4.14}
\end{equation}
We see from (\ref{4.11}) and (\ref{4.14}) that $(\partial_n G u)\sim e^{-dx}
+O(e^{-2dx})$ and $(\partial_n u G)\sim O(e^{-2dx})$.
Measure $d\mu=l^d\sinh^dxd\mu (\theta )$ induced on the boundary $\cal B$
grows as $e^{dx}$ for large $x$. Therefore we conclude that the first term
in Green's formula (\ref{6}) is negligible for large $x$
and the solution of the Laplace equation for the Dirichlet boundary problem
is indeed given by the
expression (\ref{10}). After simple computation
we get
\begin{equation}
u_D(x',\theta ')=-{1\over \Sigma_d}\int_{S^d}{f(\theta )\over \Delta^d(x', \theta , \theta ')}d\mu (\theta )
\label{4.15}
\end{equation}
for the solution and
\begin{equation}
W_D={d\over 2\Sigma_d}l^{d-1}\int\limits_{~~ S^d}\!\!\int
f(\theta ){1\over (\sin{\gamma\over 2})^{2d}}f(\theta ')d\mu (\theta )
d\mu (\theta ')
\label{4.16}
\end{equation}
for the boundary action.

Interpreting the kernel in the integral (\ref{4.16})
as a correlation function for operators ${\cal O}_D(\theta )$ consider the boundary
source term $\int_{\cal B}{\cal O}_D f d\mu$. It is finite if
the operator ${\cal O}_D\sim e^{-dx}$ for large $x$. This is an indication
that ${\cal O}_D$ should have dimension $d$.
Therefore, the correct source term should be 
$l^{d-1\over 2}\int_{\cal B}{\cal O}_D f d\mu$ and
the correlation function
for ${\cal O}_D$ reads
\begin{equation}
<{\cal O}_D(\theta ){\cal O}_D(\theta ')>
={d\over 2\Sigma_d}{1\over a^{2d}}{1\over (\sin{\gamma\over 2})^{2d}}
~~,
\label{4.17}
\end{equation}
where we introduced a scale $a$ defying measure
on the boundary $S^d$ as $a^dd\mu (\theta )$, the scale $a$ drops out
in (\ref{4.16}) and its value is completely
irrelevant.
The consideration of this section is in agreement with the results obtained 
in \cite{2}, \cite{3}.

\bigskip

{\bf 5.2 Neumann problem on $AdS_{d+1}$} \\
As in the previous subsection 
we start with the consideration of  two-dimensional case.
The Neumann condition at infinity of the space
$H^{2}$ is formulated as  follows
\begin{equation}
\left(\partial_nu-g(\phi )e^{-x}\right)|_{x\rightarrow \infty}=0~~,
\label{4.18}
\end{equation}
where $g(\phi )$ is a function on ${\cal B}=S^1$ which satisfies the condition (\ref{5}) $\int_{\cal B}g(\theta )d\mu (\phi )=0$.

We search Green's function for the Neumann problem on $H^{2}$  by analogy
with the flat case described in Section 4.6 as a solution of the inhomogeneous
Laplace equation
\begin{equation}
\Box G_N=-{1\over 2\pi l^2}
\label{4.19}
\end{equation}
satisfying the boundary condition
\begin{equation}
\partial_n G_N|_{x\rightarrow \infty}=c+O(e^{-2x})~~,
\label{4.20}
\end{equation}
where $c$ is some constant.

The solution of the equation (\ref{4.19}) can be found explicitly
\begin{equation}
G_N=-{1\over 2\pi}\ln \sinh{\sigma\over l}~~.
\label{4.21}
\end{equation}
Expanding this expression for large $\sigma$ and taking into
account (\ref{4.4}) one gets 
\begin{equation}
G_N\simeq-{1\over 2\pi}x-{1\over 2\pi}\ln \Delta +{1\over 2\pi}{e^{-2x}\over \Delta^2}~~.
\label{4.23}
\end{equation}
We see that Green's function (\ref{4.21})-(\ref{4.23})  satisfies the condition (\ref{4.20}) with $c=-{1\over 2\pi l}$.
The solution of the Neumann problem then is given by the expression
\begin{eqnarray}
&&u_N(x',\theta ')=\int_{\cal B}G_N\partial_n u \nonumber \\
&&=-{1\over 4\pi}\int_0^{2\pi}g(\phi )\ln \left(\cosh x'-\sinh x' \cos (\phi-\phi ')\right)ld\phi~~.
\label{4.24}
\end{eqnarray}
This results in the boundary action
\begin{equation}
W[u_N]=-{1\over 16\pi}\int_0^{2\pi}\int_0^{2\pi}g(\phi )\ln \sin^2({\phi-\phi '\over 2}) g(\phi ') ld\phi ld\phi '
\label{4.25}
\end{equation}
and the correlation function
\begin{equation}
K_N(\phi ,\phi ')=-{1\over 16\pi}\ln \sin^2({\phi-\phi '\over 2})
\label{4.26}
\end{equation}
what is identical (up to a constant factor) to the Neumann correlation function (\ref{3.22}) appearing
on the boundary of two-dimensional flat disk $D^2$.
It is due to the conformal equivalence of spaces $H^2$ and $D^2$.
The difference in the factor
can be removed by the rescaling the function $g(\phi )$ and/or re-defying
the radius of the circle $S^1$.

The solving the Neumann problem on higher-dimensional 
AdS spaces goes in a similar manner. 
The Neumann condition at infinity of the space $H^{d+1}$ can be imposed 
as follows
\begin{equation}
\left(\partial_n u-g(\theta )e^{-dx}\right)_{x\rightarrow \infty}=0~~,
\label{4.27}
\end{equation}
where $g(\theta )$ is a function on ${\cal B}=S^d~,~\int_{\cal B}g(\theta )d\mu (\theta )=0$. Green's function $G_N(x,x', \theta , \theta ')$
for the Neumann problem should satisfy the condition
\begin{equation}
\partial_n G_N|_{x\rightarrow \infty}=c+o(e^{-dx})~~,
\label{4.28}
\end{equation}
where $c$ is a constant and $o(z)$ is defined as $z^{-1}o(z)\rightarrow 0$
if $z\rightarrow 0$. The asymptote (\ref{4.28}) is necessary to have
$\int_{\cal B}\partial_n G_Nu=const$. Then the solution of the Neumann  problem is given by the expression (\ref{10}).
From the condition (\ref{4.28}) one finds the form of the function $G_N$ for large $x$
\begin{equation}
G_N(x,x', \theta , \theta ')=c\sigma +
o(e^{-dx})~~,
\label{4.29}
\end{equation}
where $\sigma\simeq xl+l\ln \Delta (x', \theta , \theta ')$ 
(see eq.(\ref{4.13}).
This leads us to the expression  for the  solution of the Neumann problem 
on $H^{d+1}$
\begin{equation}
u_N(x', \theta ')=cl\int_{\cal B}\ln \Delta (x',\theta ,\theta ') ~g(\theta )d\mu (\theta )+constant~~,
\label{UN}
\end{equation}
where $\Delta (x',\theta ,\theta ')$ takes the form (\ref{4.13}).
The above arguments leading to eq.(\ref{UN}) are not rigorous. We, however,
may check by direct computation that
that the function (\ref{UN}) is a solution of the Laplace equation
$\Box u_N=0$ on the whole space
$H^{d+1}$. It is seen from the fact that on  AdS space $H^{d+1}$
the function $\ln \Delta (x',\theta ,\theta ')$ is a solution of 
the inhomogeneous Laplace equation\footnote{
In the appropriate coordinate system the Laplace operator 
takes the form \\
$\Box=l^{-2}{1\over \sinh^{d}x'}\partial_{x'}\sinh^dx'\partial_{x'}
+l^{-2}{1\over \sinh^{2}x'}{1\over \sin^{d-1}\theta '}\partial_{\theta '}\sin^{d-1}\theta '
\partial_{\theta '}$.}

$$
\Box \ln \Delta (x', \theta , \theta ')={d\over l^{2}}~~.
$$
Therefore, acting the Laplace operator $\Box$ on the function  (\ref{UN})
and using the condition (\ref{5}) we find that $\Box u_N=0$. 

It is also easy to check that (\ref{UN}) satisfies the condition
(\ref{4.27}). For large $x'$ and for $\theta '\simeq \theta $ we have
for the normal derivative of (\ref{UN})
$$
l^{-1}\partial_{x'} u_N(x',\theta ')\simeq 4ce^{-dx'}\int_{\cal B}
{e^{-(2-d)x'}\over |\theta-\theta '|^2+2e^{-2x'}}g(\theta )d\mu (\theta )~~,
$$
where $|\theta-\theta '|^2=
\sum_{i=1}^d|\theta_i-\theta_i '|^2$.
It  approaches for large $x'$ the condition (\ref{4.27}) provided that we
exploit (for $k=1$)  the  
limit 
\begin{equation}
{\epsilon^{2k-d}\over (|\theta-\theta '|^2+2\epsilon^2)^k}\rightarrow \alpha_k 
\delta^d(\theta-\theta ')
\label{delta}
\end{equation}
when  $\epsilon\rightarrow 0$, where the $\delta$-function is defined by the condition $\int_{S^d}\delta^d(\theta-\theta ')d\mu(\theta )=1$ and $\alpha_k$ is
some numerical coefficient (actually, the constant $c$ in (\ref{4.28}) is determined by the condition $4c\alpha_1=1$).

Inserting the solution (\ref{UN}) into the boundary action $W[u_N]={1\over 2}
\int_{\cal B}u_N\partial_n u_N $ we obtain that
\begin{equation}
W[u_N]=c2^{-(d+1)}l^{d+1}\int_{S^d}g(\theta )~\ln \sin^2{\gamma \over 2}~
g(\theta ')d\mu (\theta )d\mu (\theta ' )~~.
\label{4.30}
\end{equation}
Thus, we find that the Neumann correlation function (defined as second derivative of $W[u_N]$ with respect to $g(\theta )$)
\begin{equation}
<{\cal O}_N(\theta ){\cal O}_N(\theta ')>=K_N(\theta ,\theta ')=c{l^{d+1}\over 2^{d+1}}\ln
\sin^2{\gamma\over 2}
\label{4.31}
\end{equation}
has universal for all AdS spaces 
logarithmic behavior. The correlation function 
(\ref{4.31}) appears to be natural if we analyze
the boundary source term $\int_{\cal B}{\cal O}_N\partial_n ud\mu$. It is finite if the operator ${\cal O}_N$ approaches finite value
for large $x$, i.e. ${\cal O}_N$ should have dimension $0$ and (\ref{4.31}) 
is the correlation function of operators of such dimension.

\bigskip

{\bf 5.3 Dirichlet and Neumann problems on 
$AdS_{d+1}$ for massive field} \\
The solution of the Laplace equation with mass (\ref{3.11}) behaves at infinity
of AdS space as follows \cite{3}
\begin{equation}
u(x,\theta )=u_0(\theta )e^{k_+x}+u_1(\theta )e^{-(k_++d)x}
~~,
\label{4.34}
\end{equation}
where $k_+=2^{-1}(\sqrt{d^2+4m^2}-d)$ and $-(k_++d)$
are two roots of the equation
$k(d+k)=m^2$. Only one of the functions $u_0(\theta )$ and $u_1(\theta )$
is independent. Fixing one of them on the boundary $S^d$ of $AdS_{d+1}$
we should be able to determine the other one.
We have the Dirichlet boundary problem when  the function
$u_0(\theta )$ is  fixed at infinity. In the other case,
fixing the function $u_1(\theta )$ we have the Neumann problem.

It is easy to see that the boundary action $W_{B}={1\over 2}\int_{\cal B}u\partial_n u$ considered on a function  of the
form (\ref{4.34}) diverges
$$
W_{B}\simeq {l^{d-1}\over 2^{d+1}}\int_{S^d}u_0^2k_+e^{(2k_++d)x}d\mu(\theta )
$$
for large $x$ and needs to be regularized. A natural way of doing this
is 
just to subtract the contribution of the leading term in (\ref{4.34})
\begin{equation}
W_{reg}=W_B[u]-W_B[u_0e^{k_+x}]~~.
\label{4.35}
\end{equation}
The regularized action then 
\begin{equation}
W_{reg}=-{dl^{d-1}\over 2^{d+1}}\int_{S^d}u_0(\theta )u_1(\theta )d\mu (\theta )
\label{4.36}
\end{equation}
is finite.
Note, that the way we define $W_{reg}$ is similar to the subtraction procedure 
one usually applies to make finite the gravitational action \cite{HH}.
One subtracts the contribution of the asymptotically
dominant part (typically it is that of flat space-time)
of the metric considering it as a background. This is exactly what we are doing
in (\ref{4.35}).

In order to calculate the regularized boundary action (\ref{4.36})
we have to find (for each type of  boundary
condition) the functional relation between $u_0(\theta )$ and $u_1(\theta )$.
We start with the considering the Dirichlet problem when $u_0(\theta )$
is fixed at infinity. The corresponding Green's function
$G_D$ is a solution of the equation $(\Box -m^2)G_D=0$ and for large $x$ behaves as
follows
\begin{equation}
G_D\simeq ce^{-(k_++d){\sigma\over l}}\simeq c{e^{-(k_++d)x}\over \Delta^{k_++d}}~~.
\label{4.37}
\end{equation}
It is important to note that the solution of the Dirichlet problem
in this case is not given by the equation (\ref{10}) since the term $(G_D\partial_nu)$ does not vanish at infinity. We have to use Green's formula (\ref{6}).
The solution then reads
\begin{equation}
u_D(x',\theta ')={cl^{d-1}\over 2^{d}}(2k_++d)\int_{S^d}{u_0(\theta)
\over \Delta^{k_++d}}d\mu (\theta )~~.
\label{4.38}
\end{equation}
The kernel ${1\over \Delta^{k_++d}}$ of the integral (\ref{4.38}) approaches
(see equation (\ref{delta})) 
$\alpha_{k_+}e^{k_+x}
\delta^d(\theta-\theta ')$ when $\gamma\rightarrow 0~(\theta \rightarrow 
\theta ')$
and is $e^{-(k_++d)x}(\sin^2{\gamma\over 2})^{-(k_++d)}$ for $\gamma\neq 0$.
The singular part of the kernel after the integration gives
the term $u_0(\theta )e^{k_+x}$ in (\ref{4.38}) what is in 
agreement with the imposed Dirichlet
condition. On the other hand, the finite part of the kernel
gives the functional relation 
\begin{equation}
u_1(\theta ')={cl^{d-1}\over 2^{d}}(2k_++d)\int_{S^d}{u_0(\theta)
\over (\sin^2{\gamma\over 2})^{k_++d}}d\mu (\theta )~~.
\label{4.39}
\end{equation}
It should be noted that the separation made in the kernel ${1\over \Delta^{k_++d}}$ on singular (when $\gamma\rightarrow 0$) and finite parts is not well-defined
since the ``finite'' part proportional to ${1\over (\sin^2{\gamma\over 2})^{k_++d}}$ diverges when $\gamma\rightarrow 0$. In fact, the finite part is defined
up to the term $a\delta^d(\theta-\theta ')$ with some (divergent)
coefficient $a$. It can be chosen to regularize the kernel
${1\over (\sin^2{\gamma\over 2})^{k_++d}}$.
Below (and earlier in (\ref{4.16}),
(\ref{4.17})) namely this definition of the kernel should be  meant.
Mathematically strict consideration of the divergent kernels
and their regularization can be found in \cite{17}, in the present
context it was done in \cite{6}, \cite{7}.

The substitution of (\ref{4.39}) into eq.(\ref{4.36}) gives us
the regularized boundary action for the Dirichlet
boundary problem
\begin{equation}
W_{reg}[u_D]=-{cdl^{2d-2}\over 2^{2d+1}}(2k_++d)\int_{S^d}u_0(\theta){1
\over (\sin^2{\gamma\over 2})^{k_++d}}u_0(\theta ')d\mu (\theta )d\mu (\theta ')~~.
\label{4.40}
\end{equation}
The boundary correlation function then reads
\begin{equation}
K_D(\theta ,\theta ')=-cdl^{2d-2} 2^{-(2d+1)}(2k_++d){1
\over (\sin^2{\gamma\over 2})^{k_++d}}~~.
\label{4.41}
\end{equation}
The mass $m$, thus, results in the anomalous dimension of the boundary 
operator ${\cal O}_{D}$. This was demonstrated in \cite{2}, \cite{3}. 
The divergence of the kernel in (\ref{4.40}), (\ref{4.41}) for $\gamma\rightarrow 0$ signals about the UV divergence of the boundary theory. Remarkably,
if we put the boundary at finite  from the center
of AdS space distance $x_B$ the 
kernel occurs to be finite what is seen from (\ref{4.38}).
The divergence appears only when $x_B$ goes to infinity.
Large $x_B$ means infra-red regime in the bulk theory. So, we find that
it is related to the UV regime of the boundary theory.  This relation 
was recently
pointed out by Susskind and Witten \cite{20}.

In the Neumann problem we fix the function $u_1(\theta )$ at the infinity of the space $H^{d+1}$. Green's formula (\ref{6}) says us that
$$
u_N=-\int_{\cal B}\left((\partial_nG_N-{k_+\over l}G_N)u_0e^{(d+k_+)x}+
(\partial_nG_N+{k_++d\over l}G_N)u_1e^{-k_+x}
\right)~~.
$$
It gives us the solution of the Neumann problem if the solution $u_N$ in
the bulk   is determined only by the function $u_1$ fixed on the boundary. It is so 
if Green's function $G_N$ satisfies
the conditions
$$
\partial_nG_N-{k_+\over l}G_N=o(e^{-(d+k_+)x})
$$
$$
\partial_nG_N+{k_++d\over l}G_N=constant~ e^{k_+x}+o(e^{k_+x})
~~.
$$
The function which satisfies these conditions is
\begin{equation}
G_N={c_1l\over k_+}e^{k_+{\sigma\over l}}+o(e^{-(d+k_+)x})~~,
\label{4.42}
\end{equation}
where $c_1$ is a constant and we include the factor ${1\over k_+}$ in order to
have the correspondence with the massless case: $G_N\simeq c_1\sigma+const$ when $k_+\rightarrow 0$.

Using (\ref{4.42}) we find 
\begin{equation}
u_N(x',\theta ')=-{c_1l^{d-1}\over  2^{d}}(2k_++d){1\over k_+}\int_{S^d}\Delta^{k_+}(x',\theta ,\theta ')u_1(\theta)d\mu (\theta )~~
\label{4.43}
\end{equation}
for the solution of the Neumann problem. 

That the function (\ref{4.43})
is a solution of the equation $(\Box-m^2)u=0$ on $H^{d+1}$ follows from the fact that
the function $\Delta^{k}(x',\theta , \theta ')$ for arbitrary $k$
satysfies the equation
\begin{equation}
\Box \Delta^{k}=k(k+d)\Delta^{k}~~.
\label{Master}
\end{equation}
Thus, for $k=k_+$ we find from (\ref{Master}) that the kernel $\Delta^{k_+}$
arising in (\ref{4.43})   
is a solution of the Laplace equation with mass $m$. Another solution
of the massive field equation corresponds to the value $k=-(k_++d)$ and 
is the kernel $\Delta^{-(k_++d)}$ (\ref{4.38}) arising in the Dirichlet problem.

If $\theta \rightarrow \theta '$ we 
have $\Delta^{k_+}(x',\theta ,\theta ')\rightarrow
e^{-(k_++d)x}\alpha_{k_+}\delta^d(\theta-\theta ')$ what is obtained from (\ref{delta}) by analytical continuation to negative $k$. The integral (\ref{4.43})
then reproduces the term $u_1(\theta )e^{-(k_++d)x}$ as it should be by the 
imposed boundary condition. The regular part of the kernel $\Delta^{k_+}$
in (\ref{4.43}) gives us the functional relation
\begin{equation}
u_0(\theta ')=-{c_1l^{d-1}\over 2^{d}}(2k_++d){1\over k_+}\int_{S^d}(\sin^2{\gamma\over 2})^{k_+} u_1(\theta)
d\mu (\theta )~~
\label{4.44}
\end{equation}
between the functions $u_0$ and $u_1$.
Substituting (\ref{4.44}) into the regularized action (\ref{4.37}) and taking the variational derivatives with respect to $u_1 (\theta )$ we find
\begin{equation}
K_N(\theta ,\theta ')=-{c_1dl^{2d-2}\over 2^{2d+1}}(2k_++d){1\over k_+}
 (\sin^2{\gamma\over 2})^{k_+}~~
\label{4.45}
\end{equation}
for the Neumann correlation function.
We see that due to the mass in the bulk the Neumann operator ${\cal O}_N$
on the boundary $S^d$ of the space $AdS_{d+1}$ acquires negative anomalous
dimension $-k_+$. As a consequence of this the Neumann correlation function
(\ref{4.45}) vanishes when the points $\theta$ and $\theta '$
coincide. However, if $2k_+$ is not integer then $n$-order derivative of
(\ref{4.45}) diverges at $\theta '=\theta$ for $n>2k_+$. This
means that $K_N(\theta , \theta ')$ (\ref{4.45}) is not
analytical at $\theta '=\theta$.  We see that mass $m$ plays the role
of regulator for the Neumann correlator (\ref{4.31}) and makes it finite 
at the coinciding points. In the bulk mass typically improves the infra-red
(IR) behavior of the theory while the short distance divergences of
correlators are of UV nature. So, what we find for the Neumann correlator seems
to be another manifestation of the relation between IR regime in the bulk and
UV regime on the boundary noted in \cite{20}.

\bigskip

\section{Hierarchy of correlators}
\setcounter{equation}0
Summarizing the results of  Sections 4 and 5 we have found that the correlators (both the Neumann and Dirichlet ones)
arising on sphere $S^d$ (considered either as a boundary 
of a domain of   flat space $R^{d+1}$
or Anti-de Sitter space $H^{d+1}$) are functions of the geodesical distance
$\gamma$ measured on $S^d$ and are constructed by means of the following
basic elements
\begin{equation}
K_0=-\ln\sin^2{\gamma\over 2}~,~~K_n={1\over \sin^n{\gamma\over 2}}~,~n>0~~.
\label{0.1}
\end{equation}
The functions (\ref{0.1}) can be defined on sphere of arbitrary
dimension  $d$. The geodesical distance $\gamma$ then is the azimuthal
angle $\theta$ between the points $P$ and $P'$ on $S^d$.
However, only on sphere of certain dimension an element (\ref{0.1})
can be considered as  a correlator of free fields.
In oder to make this statement more 
precise we consider the set of differential operators $\Delta_{n+1}=
{1\over \sin^n\theta}\partial_\theta (\sin^n\theta\partial_\theta )$.
$\Delta_{n+1}$ is the ``azimuthal'' part of the Laplace operator on $(n+1)$-dimensional sphere $S^{n+1}$. We have the following set of recurrent 
relations for the elements $K_n(\theta )$ (\ref{0.1})
\begin{eqnarray}
&&\Delta_{n+2}K_0=-{n\over 2}K_2+(n+1)
\nonumber \\
&&\left(\Delta_{n+2}-{1\over 4}m(2n-m+2)\right)K_m={1\over 4}m(m-n)K_{m+2}~~,m\neq 0
\label{0.2}
\end{eqnarray}
Note that the combination $\Delta_{n+1}-\Delta_{n}=\cot \theta \partial_\theta$
is the first oder differential operator. The elements (\ref{0.1}) satisfy also the first oder differential recurrent relations
\begin{eqnarray}
&&\cot \theta \partial_\theta K_0=-{1\over 4}K_2+{1\over 2} \nonumber \\
&&(\cot \theta \partial_\theta-{n\over 2})K_n=-{n\over 4}K_{n+2}~~,
\label{0.3}
\end{eqnarray}
which are  analogous to the relations (\ref{2.5}) for the
correlators arising on plane boundary.

We will say that $K$ is primary correlator on $S^d$ if $K\sim <\phi (P)
\phi (P')>$ where $\phi (P)$ is a  field on $S^d$ satisfying free
(second order) field equation. In other words, the primary correlator
is the one on which the recurrent sequence (\ref{0.2}) stops
for certain $n$ and $m$
and  does not produce a new correlator. 
Non-primary correlators are correlators of composite operators
built from the free fields.
We say that correlator $K$ is a descendant
 if it is obtained by differentiating  a primary
correlator. One can see from the relations (\ref{0.2}) that
a correlator $K_n$ is primary on $(n+2)$-dimensional  sphere $S^{n+2}$.
Indeed, we find from (\ref{0.2}) that 
\begin{eqnarray}
&&\Delta_2 K_0=1 \nonumber \\
&&\left(\Delta_{n+2}-{n(n+2)\over 4}\right) K_n=0~~.
\label{0.4}
\end{eqnarray}
It is easy to recognize the conformal nature of the operators
in (\ref{0.4}).
Therefore, $K_n$ is correlator of conformal field $\phi (x)$ on sphere
$S^{n+2}$ satisfying the conformal field equation
\begin{eqnarray}
&&\Delta_2 \phi=c~{\cal R}~,n=0 \nonumber \\
&&\left(\Delta_{n+2}-{n\over 4 (n+1) }{\cal R}\right) \phi=0~,n>0
\label{0.5}
\end{eqnarray}
where $c$ is two-dimensional central charge and the scalar 
curvature ${\cal R}$ of $(n+2)$-dimensional unit sphere $S^{n+2}$
is $(n+2)(n+1)$.  There are two different
points of view on the correlation functions under consideration.
For example, the Neumann correlator (\ref{3.27}) of the boundary operators
${\cal O}_N$ can be viewed as entirely
arisen on sphere $S^d$ with only 
component $K_{d-2}$ being a primary correlator.
In the dual picture the operator ${\cal O}_N$ is considered to be 
a composite of the free field operators 
${\cal O}[S^m]$ having support on sphere $S^m$. 
 The correlator (\ref{3.27}) arises then for points $P$ and $P'$ lying on the intersection
of the set of spheres $S^{d+1}$, $S^{d}$, ..., $S^{d-m}$, ..., $S^2$ with
a component $K_{d-m-2}$  being the primary
correlator on sphere $S^{d-m}$. 

As another illustration of this let us consider 
the correlator $K_0$. It arises as the Neumann correlator on the circle
$S^1$ being boundary of two-dimensional disk $D^2$. On the other hand, $S^1$ can be
considered as a meridian on 2d sphere $S^2$ and $K_0$ is the primary
correlator of conformal scalar field on $S^2$ described by the action
\begin{equation}
W[S^2]=\int_{S^2}\left({1\over 2}(\nabla\phi )^2-cR\phi \right)~~.
\label{0.6}
\end{equation}
The correlator $K_2$ arises on $S^1$ in the Dirichlet problem on
$D^2$. On $S^2$ it is interpreted as a descendant of $K_0$ since we have from
(\ref{0.2}) that $-2\partial^2_\theta K_0(\theta )=K_2(\theta )$. 
On the other hand, $K_2(\theta )$ is the primary conformal correlator 
on four-dimensional sphere $S^4$.

As we saw in the previous Section, the correlator $K_0$ also
arises on sphere $S^d$  in the Neumann
problem on Anti-de Sitter space $AdS_{d+1}$  
and the present analysis shows that
it has natural two-dimensional description.
The Dirichlet problem on $AdS_{d+1}$ produces the correlator
$K_{2d}$ on the boundary $S^d$. It is primary on $(2d+2)$-dimensional
space $S^{2d+2}$. 
In the context of the works \cite{1},
\cite{2}, \cite{3} the five-dimensional AdS space is of interest.
The Dirichlet correlator arising on  $4$-dimensional boundary $S^4$
is $K_8$. On $S^4$ it
is a descendant of the primary correlator $K_2$.
At the same time it has a dual $10$-dimensional description on $S^{10}$
as a primary conformal field correlator.

\bigskip

\section{Phases on the boundary of AdS space}
\setcounter{equation}0
In this Section we want to point out on the existence of various phases of the
field theory living on the boundary of Anti-de Sitter space.
These phases may arise in essentially two different ways.

\medskip

{\bf 7.1 Neumann and Dirichlet phases} \\
In the first way phases arise on the boundary of AdS space when we impose 
the mixed 
boundary condition similar to what we considered in  Section 2
for  flat space. This condition should be consistent with the asymptotic 
behavior (\ref{4.34}) of a solution of the massive Laplace equation
in AdS space. In order to formulate it we introduce the field
$\tilde{u}=ue^{-k_+x}$ with the asymptotic behavior
$$
\tilde{u}\simeq u_0(\theta )+u_1(\theta )e^{-(2k_++d)}~~
$$
at infinity.
In terms of the field $\tilde{u}$ the third type boundary condition can be 
imposed as follows
\begin{equation}
\partial_n \tilde{u}+he^{-(2k_++d)x}\tilde{u}=g(\theta )e^{-(2k_++d)x}
\label{5.1}
\end{equation}
or, equivalently,
$$
hu_0(\theta )-(2k_++d)l^{-1}u_1(\theta )=g(\theta )~~.
$$
As it is seen from (\ref{5.1}) the  coupling $h$ being considered
as a function on the boundary
has dimension $(2k_++d)$. The  dimensional analysis occurs to be  
 useful in finding the 
asymptotic behavior of the corresponding correlation function $K_{(h)}$
arising on the boundary. We have
\begin{equation}
K_{(h)}\simeq {1\over k_+}\sigma^{k_+}+h\sigma^{4k_++d}+O(h^2)
\label{5.2}
\end{equation}
for small $h$ and 
\begin{equation}
K_{(h)}\simeq {1\over h^2}{1\over \sigma^{2d+2k_+}}+O({1\over h^3})~~,
\label{5.3}
\end{equation}
for large $h$,
where $\sigma=a\sin{\gamma\over 2}$ and we omitted the numerical
coefficients. The equations (\ref{5.2})-(\ref{5.3}) mean that
$K_{(h)}\simeq K_N$ for small $h$ and $K_{(h)}\simeq {1\over h^2}K_D$
for large $h$ in agreement with the consideration of  Section 2.
In the massless case $(k_+\rightarrow 0)$ the term ${1\over k_+}\sigma^{k_+}$
in (\ref{5.2}) should be replaced by $\ln \sigma$.  The picture of phases
arising on the boundary of AdS space
is similar to that we had for  flat space in Section 3.
The only essential difference is that the coupling $h$ in AdS space
has higher scaling dimension $(2k_++d)$ compared to the flat
space (where $h$ has dimension $1$). For a finite value $h$
there are two phases of the boundary system. In the regime
$(h\sigma^{d+2k_+})<<1$ the free (Neumann) phase is restored
and the correlation function behaves as (\ref{5.2}).
On the other hand, for $(h\sigma^{d+2k_+})>>1$ we have the Dirichlet 
(strong coupling) phase. 
Thus, the coupling $h$ brings some natural scale on the boundary so that the boundary theory at a finite $h$ is not conformal, though, it interpolates between
the conformal phases.

It should be noted that the Neumann phase arising on the boundary
of AdS space is quite different from what we had in flat space.
In the massive case the Neumann correlator (\ref{5.2})  
does not diverge at the coinciding
points. Though, for arbitrary mass ($k_+$) it is non-analytical 
at the point $\sigma=0$ in the sense that higher order derivatives of $K_N(\sigma)$ become divergent there. When mass is zero ($k_+=0$) the Neumann correlator
 $K_N\sim \ln \sigma$ is logarithmic. Remarkably,
it is so in all AdS spaces independently of the dimension.

We stress that the correlators arising in the Neumann phase well behave at the coinciding points. In the massless case the divergence  is logarithmic there
and, therefore, is integrable. So, integrals involving $K_N$ are finite.
In the massive case the situation even better since the Neumann correlator
is out of divergence at all. In the Dirichlet phase the correlators diverge as
$\sigma^{-2d}$ in massless case and as $\sigma^{-2(d+k_+)}$ in massive case.
These divergences are not integrable. Therefore, we need to use some
regularization in order to give sense to the integrals involving
$K_D$. Tracing these divergences to the UV behavior of the boundary theory
one could say that the theory is finite in the Neumann phase and renormalizable in the Dirichlet phase. Note, that the presence of mass in the bulk improves
the behavior of the Neumann correlator and makes worse the behavior of the 
Dirichlet one.

The Neumann correlator arising at the infinity of AdS space takes  the same
form as the correlator arising in the Neumann problem on the flat
disk $D^2$ with boundary $S^1$. The circle $S^1$ can be considered as
lying
on the sphere $S^d$ (boundary of AdS space) and joining
the points $P$ and $P'$. The geodesical distance 
$\gamma$ between the points $P$ and $P'$ then is  the angle $|\phi-\phi '|$
measured on the circle $S^1$. 
Two bulk theories,  one is living on $AdS_{d+1}$ and another on $D^2$,
produce the same
correlator on the circle. 
On the other hand, the corresponding boundary theories (living on $S^1$ and $S^d$ respectively) both have dual description as
a conformal theory (\ref{0.6}) on two-dimensional sphere $S^2$.
This fact, in particular, 
proposes that the 
Neumann phase on the boundary of AdS space may have infinite-dimensional  underlying
conformal group
of symmetry. It is in contrast with the Dirichlet phase whose conformal symmetry
is finite-dimensional.

In the construction considered in \cite{1}, \cite{2}, \cite{3}
the Anti-de Sitter space is $AdS_{5}$ and the four-dimensional
theory on the boundary is identified with the large $N$ limit of a
 super Yang-Mill
theory with gauge group $SU(N)$. The scalar Dirichlet boundary operator then should have
dimension $4$ and can be constructed from the Yang-Mills field $A_{\mu}$:
${\cal O}_D=Tr(F_{\mu\nu}F^{\mu\nu})$. In the Neumann phase the boundary 
operator should have dimension $0$. It seems that no such operator can be constructed from local combinations of $A_\mu$ and its derivatives.
A possible candidate for the boundary operator in the Neumann phase
 is the non-local one
$$
{\cal O}_N[P]\sim Tr{\cal P} \exp\int_{C_P}A_\mu dx^\mu ~~,
$$
where $C_P$ is a circle shrinking to the point $P$.
It is also an interesting question as what meaning in terms of the Yang-Mills
theory may have the boundary coupling $h$.

\bigskip

{\bf 7.2 Infinity-horizon transition in AdS like space} \\
In general, the location of the boundary in the AdS space can be arbitrary.
One may put it at $x=x_B$ and consider the domain $0\leq x \leq x_B$
of the AdS space. When the boundary
moves across AdS space this induces some automorphism in
the theory living on the boundary. In some cases this automorphism
is trivial and the boundary field theory does not depend  on
the choice of the boundary. This is the case in three dimensions
when the theory in the bulk is the Chern-Simons theory. It induces
a conformal (Wess-Zumino) theory on the boundary which remains the same
no matter where one chooses the boundary. Note, that this is a feature of
this concrete model which is mainly due to the
absence in the bulk of the propagating degrees of freedom
in the Chern-Simons theory. In general,  the moving  the boundary 
may lead to drastic deformation of the field theory on the boundary.
In the case under consideration, the theory in the bulk is described by the Laplace equation. Varying $x_B$, the location of the boundary,
we find that there are two limiting cases:
when the boundary $\cal B$ approaches infinity ($x_B\rightarrow \infty~,
~~{\cal B}={\cal B}_\infty$) and when it shrinks ($x_B\rightarrow 0~,
~~{\cal B}={\cal B}_0$) to the center of the AdS space, in both cases the boundary is topologically sphere. In the later case the domain $0\leq x \leq x_B$
is well approximated by  domain $0\leq r\leq R$
of  flat space $R^{d+1}$ with sphere
$S^d$ of radius $R=lx_B$ as a boundary. The classical field theory
on such domain and the correlation functions arising on its boundary
$S^d$ were considered in Section 4. The theory on the boundary $\cal B$
of the AdS space
is completely characterized by the scaling dimensions of the operators
${\cal O}_D$ (in the Dirichlet problem) and ${\cal O}_N$ (in the
Neumann problem).
When the boundary is at infinity the dimension of ${\cal O}_N$ is $-k_+$
(in the massless case it is $0$) while the dimension of 
${\cal O}_D$ is $2(d+k_+)$. On the opposite end,  when
the boundary stays arbitrary close to the center ($x=0$) of the AdS space
we have dimensions $(d-1)$ and $(d+1)$ respectively for the operators
${\cal O}_N$ and ${\cal O}_D$. 
The limit $x_B\rightarrow 0$ is correctly accomplished if we 
deal with the rescaled operators  $\tilde{\cal O}_N$ and 
$\tilde{\cal O}_D$ defined as $\tilde{\cal O}_N=({x_B\over a})^{-{(d-1)\over 2}}{\cal O}_N$ and $\tilde{\cal O}_D=({x_B\over a})^{-{(d+1)\over 2}}{\cal O}_D$ 
where $a$ is an arbitrary (finite) scale parameter playing the role
of the radius of the sphere $S^d$. One then finds that 
the scaling dimensions  of the operators
$\tilde{\cal O}_N$ and $\tilde{\cal O}_D$
are independent of the mass $m$ and the 
correlation functions are effectively massless.
The Neumann correlation function is presumably given by the series (\ref{3.27})
with respect to the variable $\sigma=2a\sin{\gamma\over 2}$.
It includes the logarithmic term $\ln\sigma$. It happens that only this term
survives (in the massless case) when the boundary moves to
the infinity where the Neumann correlator (\ref{4.31})
has universal logarithmic behavior.

The theory arising on ${\cal B}_\infty$ is not much sensitive
to deformations made deep inside
the AdS instanton. It is not so for the theory living
on the boundary $ {\cal B}_0$ because the topology of 
$ {\cal B}_0$ may change drastically under some
deformations. An interesting and important way to deform
the AdS space is to put a black hole inside.
The complete manifold  remains asymptotically AdS space. The topology
of the manifold near the center, however, is modified
by the presence of the black hole horizon.
It is  that of the product space $D^2\times \Sigma$,
where $D^2$ is two-dimensional disk and 
$\Sigma=S^{d-1}$ is the horizon surface.
So that the boundary ${\cal B}_0^h$ is no more a sphere $S^d$
but the product space ${\cal B}_0^h=S^1\times \Sigma$. 
According to (\ref{3.18}) and (\ref{3.28}) the correlation functions arising on
${\cal B}_0^h$ factorize $K_{{\cal B}_0^h}=\delta_\Sigma~ K_{S^1}$,
where $\delta_\Sigma$ is delta-function on $\Sigma$ and
$K_{S^1}$ is the circle correlation function arising on $S^1$
considered as a boundary of  two-dimensional disk $D^2$.
The averaged correlation function obtained by integration over $\Sigma$,
$\int\limits_{~~\Sigma}\!\!\int K_{{\cal B}_0^h}=A_\Sigma K_{S^1}$,
is proportional to the horizon area $A_\Sigma$.
On ${\cal B}_0^h$ 
the boundary operators $\tilde{\cal O}_N$ and $\tilde{\cal O}_D$
 have the effective scaling dimensions $0$ and $1$ respectively.
The Neumann correlator then 
\begin{equation}
K_{S^1}=K_N(\phi ,\phi ')=-{1\over 4\pi}\ln\sin^2{\phi-\phi '\over 2}
\label{5.4}
\end{equation}
is purely logarithmic
and is identical (after identifying the circle $S^1$ with geodesic
joining the points $P$ and $P'$
on ${\cal B}_\infty=S^d$) to 
the Neumann correlator on ${\cal B}_\infty$. So, in the case of the
black hole the Neumann phase remains unperturbed in the 
transition from the infinity to the horizon.
As we discussed previously, the correlation function $K_N(\phi ,\phi ')$
has natural two-dimensional description as a correlator
of free conformal  fields on sphere $S^2$
(the circle $S^1$ is considered as a meridian on $S^2$).
Therefore, the Neumann phase arising on the boundary ${\cal B}_0^h$ has dual 
conformal field theory description on the space $S^2$.
On the other hand, the correlation function
\begin{equation}
K_{S^1}=K_D(\phi , \phi ')=-{1\over 8\pi a^2\sin^2({\phi-\phi '\over 2})}
\label{5.5}
\end{equation}
arising on ${\cal B}_0^h$ in the Dirichlet phase 
has a description as a primary correlator  of 
conformal field theory on
$4$-dimensional sphere $S^4$. So, in the Dirichlet case
the boundary field theory on
${\cal B}_0^h$ is dual to the conformal field theory on $S^4$.
It is interesting to note that in four dimensions the correlator
of the form (\ref{5.5}) appears also as a bulk correlation function
of free quantum fields $<\varphi (P) \varphi (P')>$ when both 
$P$ and $P'$ lie on the horizon $\Sigma=S^2$, $\theta$ in this case is the azimuthal angle on $\Sigma$. The bulk correlation function in this case is
Green's function calculated in the Hartle-Hawking state. 
When at least one of the points lies on the horizon
it can be found  exactly  as was demonstrated by Frolov \cite{21} long ago
for a charged rotating black hole. It would be nice to understand
better the coincidence of two correlation functions.

We see, thus, that the Dirichlet phase on the boundary
deforms
essentially in the transition from infinity to the horizon
in AdS like space. This manifests in the scaling dimension
of the Dirichlet boundary operator (that changes from
value $d$  at infinity to $1$ at the horizon) and in the behavior
of the corresponding correlation function. The Neumann phase
is, however, more stable under this transition. The scaling dimension
$0$ of the Neumann boundary operator  and the logarithmic behavior
of the correlator (it is so up to the factor $\delta_\Sigma$ at
the horizon) remain the same. The boundary near black hole horizon,
${\cal B}_0^h=S^1\times\Sigma$, has structure typical for the Euclidean
description of a thermal field theory (the circumference of $S^1$ being
the inverse temperature). Therefore,
in terms of the boundary theory
this transition seems to correspond to the high temperature
regime. However, a more detail investigation is required.

\section*{Acknowledgements}
I have benefited from discussions with Valery Frolov.
This work was supported in part by the Natural Sciences and 
Engineering Research Council
of Canada.

\end{document}